\documentclass[12pt, draftclsnofoot, onecolumn]{IEEEtran}
\usepackage{blindtext, graphicx}
\usepackage {tikz}
\usepackage{caption}
\let\labelindent\relax
\usepackage{amsmath}
\usepackage{amssymb}
\usepackage{algorithm}
\usepackage{algpseudocode}
\usetikzlibrary {positioning}
\usepackage{tkz-euclide}
\usepackage{algorithm,algcompatible,amsmath}
\usetkzobj{all}
\usepackage[T1]{fontenc}
\usepackage[utf8x]{inputenc}
\usetikzlibrary{calc}
\usepackage{amsthm}
\usepackage{hyperref}
\usepackage{amsmath}
\usepackage{thmtools}

\definecolor {processblue}{cmyk}{0.96,0,0,0}
\usepackage{enumitem}
\usepackage{balance}
\usepackage{placeins}
\usepackage{algorithm}
\usepackage{algpseudocode}
\usetikzlibrary {positioning}
\usepackage{tkz-euclide}
\usepackage{algorithm,algcompatible,amsmath}
\usetkzobj{all}
\usepackage[T1]{fontenc}
\usetikzlibrary{calc}
\usepackage{amsthm}
\usepackage{hyperref}
\usepackage{amsmath}
\usepackage{thmtools}
\usepackage{verbatim}

\usepackage{subcaption}
\newcommand\tab[1][1cm]{\hspace*{#1}}
\usepackage[font={footnotesize}]{caption}
\usepackage{caption} 
\usepackage{stfloats}
\usepackage{lipsum}

\usepackage[ruled,vlined,linesnumbered,algo2e]{algorithm2e}
\usepackage{bbm}
\usepackage{amsmath,amssymb}

\DeclareMathOperator*{\argmax}{arg\!\,max}

\makeatletter
\def\ps@IEEEtitlepagestyle{
  \def\@oddfoot{\mycopyrightnotice}
  \def\@evenfoot{}
}
\def\mycopyrightnotice{
  {\footnotesize
  \begin{minipage}{\textwidth}
  \centering
  This work is to be submitted to the IEEE for possible publication.  Copyright may be transferred without notice, after which this\\ version may no longer be accessible.
  \end{minipage}
  }
}

\ifCLASSINFOpdf
 
\else
 
\fi

\begin{document}

\title{Nearly Optimal Scheduling of Wireless Ad Hoc Networks in Polynomial Time\vspace{0.5cm}}

\author{
        \IEEEauthorblockN{\normalsize
        Alper Köse\IEEEauthorrefmark{1}\IEEEauthorrefmark{3},
        Noyan Evirgen\IEEEauthorrefmark{5},
        Hakan Gökcesu\IEEEauthorrefmark{4},
        Kaan Gökcesu\IEEEauthorrefmark{2},
        Muriel Médard\IEEEauthorrefmark{1}\IEEEauthorrefmark{2}}
        \IEEEauthorblockA{
                \small\\
                \vspace{0.4cm}
                \IEEEauthorrefmark{1} Research Laboratory of Electronics, Massachusetts Institute of Technology\\
                \vspace{-0.075cm}
                \IEEEauthorrefmark{2} Department of Electrical Engineering and Computer Science, Massachusetts Institute of Technology\\
                \vspace{-0.075cm}
                \IEEEauthorrefmark{3} Department of Electrical Engineering, École Polytechnique Fédérale de Lausanne\\ 
                \vspace{-0.075cm}
                \IEEEauthorrefmark{4} School of Computer and Communication Sciences, École Polytechnique Fédérale de Lausanne\\
                \vspace{-0.075cm}
                \IEEEauthorrefmark{5} Department of Information Technology and Electrical Engineering, ETH Zürich\\
                \vspace{0.1cm}
%                \{akose, gokcesu, medard\}@mit.edu, nevirgen@ethz.edu, hakan.gokcesu@epfl.ch
                \vspace{-0.5cm}
        }
    
}
\IEEEoverridecommandlockouts

\maketitle

\begin{abstract}

In this paper, we address the scheduling problem in wireless ad hoc networks by exploiting the computational advantage that comes when such scheduling problems can be represented by claw-free conflict graphs where we consider a wireless broadcast medium. It is possible to formulate a scheduling problem of network coded flows as finding maximum weighted independent set (MWIS) in the conflict graph of the network. Finding MWIS of a general graph is NP-hard leading to an NP-hard complexity of scheduling. In a claw-free conflict graph, MWIS can be found in polynomial time leading to a throughput-optimal scheduling. We show that the conflict graph of certain wireless ad hoc networks are claw-free. In order to obtain claw-free conflict graphs in general networks, we suggest introducing additional conflicts (edges) while keeping the decrease in MWIS size minimal. To this end, we introduce an iterative optimization problem to decide where to introduce edges and investigate its efficient implementation. Besides, we exemplify some physical modifications to manipulate the conflict graph of a network and also propose a mixed scheduling strategy for specific networks. We conclude that claw breaking method by adding extra edges can perform nearly optimal under the necessary assumptions.

\end{abstract}


\IEEEpeerreviewmaketitle

\section{Introduction}

We study the scheduling problem in wireless ad hoc networks. In a wireless broadcast medium, networks are usually interference limited and hence, interfering transmissions cannot be done simultaneously. On the other hand, it is necessary to maximize the number of simultaneous transmissions in order to obtain a high throughput in the network. This trade-off enforces us to do scheduling which aims to maximize the number of non-interfering simultaneous transmissions in considered time slot.

Arikan \cite{arikan1984some} proves that scheduling problem is NP-complete for packet radio networks which is the earliest version of wireless networks. Ephremides and Truong \cite{ephremides1990scheduling} study the problem of scheduling broadcast transmissions in a multihop interference limited wireless network while aiming to optimize throughput. They show that the problem is NP-complete. Sharma et al. \cite{sharma2006complexity} also consider the problem of throughput optimal scheduling in wireless networks subject to interference constraints where they assume no two links within K hops can successfully transmit in the same time slot. They conclude that the problem can be solved in polynomial time for K=1 whereas it is NP-hard for K>1. Hajek and Sasaki \cite{hajek1988link} give polynomial time algorithms for link scheduling in a spread spectrum wireless network where each node is allowed to converse with only one other node at a time. Our modeling of possible transmissions in interference limited network setup and approach using conflict graphs are same with Traskov et al.'s \cite{traskov2012scheduling} work. Therefore, in our case, scheduling has an NP-hard complexity as in \cite{traskov2012scheduling} for general conflict graphs.

Due to the complexity of the scheduling problem, common approach is to propose an approximate solution. For example, Traskov et al. \cite{traskov2012scheduling} propose an approach that greedily chooses maximal independent sets instead of finding maximum independent set since the complexity of the latter is NP-hard despite giving the optimal solution. Another example is Bao et al. \cite{bao2001new} who propose a suboptimal interference scheme where two nodes within two hops cannot transmit simultaneously. As known, there is no optimal solution for the scheduling problem in polynomial time and in this work, we propose to change perspective and investigate this problem from another angle as explained in the following paragraphs.

According to our assumptions and Protocol model that Gupta and Kumar \cite{gupta2000capacity} defines, we construct the conflict graph of a given network where we model possible transmissions as the vertices of the conflict graph. We define the neighbors of a transceiver as the set of transceivers this transceiver can transmit. To model possible transmissions, we first find the neighbors of each transceiver, then we assign each transceiver as the sender and every possible combination of its neighbors as its possible receivers, implicitly meaning that there is a directed hyperedge from sender to its possible receivers in the network setup. In the end, the number of possible transmissions for a transceiver is equal to the number of subsets of its set of neighbors except the zero set. An edge between two vertices of a conflict graph means that it is not possible to schedule these two transmissions for the same time slot since they interfere with each other. To find the edges between modeled vertices in the conflict graph, again we use our assumptions with the Protocol model and give the interference relationships between possible transmissions. Note that our setup is different than link based scheduling due to broadcast modeling approach.

Network coding concept is introduced by Ahlswede et al. \cite{ahlswede2000network} and can be used to improve the performance of networks. Ho et al. \cite{ho2003randomized,ho2006random} study the random linear network coding approach and show that it can achieve capacity in multisource multicast networks. We implicitly consider random linear network coding over considered wireless network in a bandwidth limited regime as in \cite{traskov2012scheduling} and thus our conflict graph construction accounts for this. The ultimate aim is to compute an optimal network coding subgraph and a schedule that can support it. We require network coding subgraph to lie in the independent set polytope of the conflict graph so that the subgraph can be partitioned into a combination of valid schedules. Although this optimization has an NP-hard complexity in general, it can be done in polynomial time for claw-free conflict graphs. So, conflict graph contains the combinatorial difficulty of the scheduling problem. In this work, we concentrate on the scheduling problem and consider the graph-theoretical side to get claw-freeness in networks.

Scheduling can be modeled as maximum weighted independent set (MWIS) problem in the conflict graph \cite{traskov2012scheduling,tassiulas1992stability,sundararajan2006systematic,sundararajan2007network}. Therefore, scheduling complexity is equivalent to the complexity of finding MWIS in the derived conflict graph of given wireless network as shown by Traskov et al. \cite{traskov2012scheduling}. Since there are algorithms \cite{minty1980maximal,nakamura2001revision,schrijver2003combinatorial,faenza2014solving} that can find MWIS in polynomial time in claw-free graphs, we can do polynomial time scheduling if we can get a claw-free conflict graph.

We investigate some families of networks, which have claw-free conflict graphs, including line networks and tree networks. Since there are many limitations to construct networks which have claw-free conflict graphs, we are able to set up such networks under very specific assumptions. Typical wireless networks usually do not have claw-free conflict graphs, thus we propose to add conflict edges to their conflict graphs or to do minor and necessary modifications in networks to reach claw-freeness in their conflict graphs. Note that, introducing only a few edges can break all claws and hence give a nearly optimal performance in many cases and this is confirmed with our simulations. Another advantage of this method is that we are able to automatically decide between which nodes the edges must be introduced. Thus, we are able to incorporate an optimization problem which breaks all the claws by adding edges between nodes so that the decrease in the optimal scheduling throughput, i.e. the weighted size of the MWIS, is minimal. Second possible method, physical modifications in network, requires network flexibility. Also, there must be an autonomous system to immediately propose the modifications that should be done. Lastly, we propose a heuristic mixed scheduling algorithm for the networks in which all claws in the conflict graph come from a specific part of the network. Using this approach, we are able to do throughput optimal scheduling for the claw-free part of the network whereas we use approximate scheduling for the rest and combine the solutions in the end. This paper is an extension of \cite{2017arXiv171101620K}.

In short, our contributions are the following:
\begin{itemize}

\item Introducing some families of networks which can be scheduled in polynomial time;

\item Adding new edges to conflict graph with minimal decrease in MWIS' total weight and without any intervention to network setup in order to make network suitable for polynomial time scheduling;

\item Suggesting physical modifications in the network setup to make network suitable for polynomial time scheduling;

\item Introducing a novel heuristic mixed scheduling algorithm based on the location distribution of claws in physical network.

\end{itemize}

The remainder of this paper is organized as follows. In Section \ref{cons}, we detail on the conflict graph construction and present different scenarios, which are on line, tree and diamond networks for which the conflict graphs are claw-free. In Section \ref{sec:model2}, we explain the methods that can be used in order to make a general network suitable for polynomial time scheduling. In Section \ref{breakclaw}, we detail on the claw-breaking strategy that we use in the conflict graph. In Section \ref{subsec:phys}, we exemplify the possible physical modifications that can be done in the network to get claw-freeness in the conflict graph. In Section \ref{mixedsc}, we propose a heuristic mixed scheduling algorithm to exploit the advantage of a partition of a network when claws, in network's conflict graph, come from a specific part of network. We present the simulation results and evaluate the near-optimality of the claw breaking strategy in Section \ref{simulasyon}. We conclude the paper in Section \ref{concl}. 

\section{Constructions for Polynomial Time Scheduling}
\label{cons}

\textbf{Definition 1 - Claw-free graph}: \textit{A graph $\mathcal{G}=(\mathcal{V}, \mathcal{E})$ is claw-free if none of its vertices $\mathcal{V}$ has three pairwise nonadjacent neighbours \cite{chudnovsky2005structure}.}

\textbf{Definition 2 - Independent Set}: \textit{Given an undirected graph $\mathcal{G}=(\mathcal{V}, \mathcal{E})$, a subset of vertices $\mathcal{S}\subseteq \mathcal{V}$ is an independent set if $\{i,j\}\notin \mathcal{E}$ is satisfied for all $i$ and $j$ in $\mathcal{S}$.}\\

\vspace{-0.5cm}

Throughout the scenarios, we use the Protocol model \cite{gupta2000capacity} and K-hop interference model \cite{sharma2006complexity} with small variations to represent networks instead of the Physical model \cite{gupta2000capacity}, which takes $SINR$ levels into account.

\subsection{Scenario I - Line Networks}
\label{sc1}

We have a wireless ad hoc network with $n$ transceivers with the following assumptions:

\begin{itemize}
\item A transceiver can receive from at most one transceiver in a time slot.
\item Omnidirectional antennas are deployed.
\item Time division duplex transceivers are used.
\end{itemize}

We model interference between transmissions with the Protocol model. Assume transceiver $i_a$ is transmitting to transceiver $i_b$ while transceiver $i_c$ is transmitting to another one. Then, the transmission between $i_a$ and $i_b$ will be successful if and only if following inequalities are satisfied:
\begin{equation}
\label{eq:1}
|P_{i_a}-P_{i_b}|\leq r_{T}
\end{equation}

\vspace{-0.5cm}

\begin{equation}
\label{eq:2}
|P_{i_b}-P_{i_c}|\geq (1+\Delta)|P_{i_a}-P_{i_b}|
\end{equation}

where $P_{i_a}$ is the position of transceiver $i_a$. Inequality (\ref{eq:1}) means that transceivers have a maximum range of transmission $r_T$. Inequality (\ref{eq:2}) means that in a communication pair, among all transmitting nodes, the receiver of this pair must be closest to its transmitter with a guard zone $\Delta$. Without loss of generality $\Delta > 0$ and can be chosen arbitrarily small for simplicity. 

Conflict graph, as the name suggests, is the graph of transmissions which conflicts with each other. We need to represent the potential conflicts among the transmissions since they are directly related with scheduling. In a conflict graph, transmissions are represented by vertices and conflicts are represented by edges. A conflict exists under certain conditions which depend on the assumptions on network. After identifying all of the potential conflicts among the transmissions, it is possible to check if the conflict graph is claw free. 

We use a similar approach for the construction of conflict graph $\mathcal{G}=(\mathcal{V}, \mathcal{E})$ in a wireless network with Traskov et al. \cite{traskov2012scheduling}, where $\mathcal{V}$ is the set of possible transmissions and $\mathcal{E}$ is the set of conflicts. Vertices and edges of the conflict graph are found as follows. Let us assume $\mathcal{T}$ is the set of transceivers, $\{i_1,i_2,...,i_n\}\in \mathcal{T}$. There is a set $N(i_x)$ for $\forall i_x \in \mathcal{T}$, whose elements are neighbors of $i_x$. Then, we are able to define the set of possible receivers. There is a set $Y_x = P(N(i_x))$ for $\forall i_x \in \mathcal{T}$ if $|N(i_x)|\geq 1$ where $P(\cdot)$ is the power set of $\{\cdot\}$ without the empty set. We find the possible transmission by defining $i_x$ as the sender and each element of $Y_x$ as the receiver set, respectively, and we do this for $\forall i_x \in \mathcal{T}$. In the end, we can symbolize possible transmissions as $S_k = (i_k,J_k)$ for $k=\{1,...,m\}$ where $m=\sum_{t=1}^{n} |Y_t|$. Say $S_{1},S_{2}\in \mathcal{V}$, then $\{S_{1},S_{2}\}\in \mathcal{E}$ if any of the following conditions hold for $S_{1} = (i_1,J_1)$ and $S_{2} = (i_2,J_2)$, which means they cannot be scheduled for the same time slot:

\begin{enumerate}[label=\textbf{C2.1.\arabic*},ref=C2.1.\arabic*,labelindent=20pt,itemindent=1em,leftmargin=25pt]
\item \label{c1} $i_{1}=i_{2}$.
\item \label{c2} $(i_{1}\in J_{2}) || (i_{2}\in J_{1})$.
\item \label{c3} $J_{1}\cap J_{2}\neq \emptyset$.
\item \label{c4} $|i_{2}-j|\leq (1+\Delta)|i_{1}-j|$ for $\exists j\in J_{1}$.
\item \label{c5} $|i_{1}-j|\leq (1+\Delta)|i_{2}-j|$ for $\exists j\in J_{2}$. \label{last-item}
\end{enumerate}

In the network, the condition $|N(i_x)|\leq K$ must be satisfied for $\forall i_x \in \mathcal{T}$ in order to complete the conflict graph setup in a reasonable time. $|N(i_x)|$ is the number of neighbors that transceiver $i_x$ has as we defined earlier and $K$ is a small integer that can arbitrarily be specified. For example, a reasonable assumption is that $K=5$. Computational complexity of creating vertices to model possible transmissions becomes $O(n2^{K})$ since we have $n$ transceivers in the network and each transceiver can lead to at most $2^{K}-1$ possible transmissions in the conflict graph. Then, complexity of adding necessary edges between vertices to model conflicts is $O(n^{2}2^{2K})$. In overall, they imply a polynomial time complexity $O(n^{2})$ if we satisfy $|N(i_x)|\leq K$ for $\forall i_x \in \mathcal{T}$, $K$ being a small integer. Think of a scenario where this condition is not satisfied. Let us have a network with $n$ nodes and each node is in the transmission range of all other nodes. In this case, we have to set $n(2^{n-1}-1)$ vertices in the conflict graph to be able to represent all possible transmissions, meaning that we face an exponential complexity. Modeling edges is even more computationally complex, therefore this is not computationally feasible, even when there is small number of transceivers such as $n=20$. This is why we have to set a bound for the number of neighbors of transceivers. We could say that $|N(i_x)|\leq \log n$ in general, but this would be too restrictive for networks having low number of transceivers.

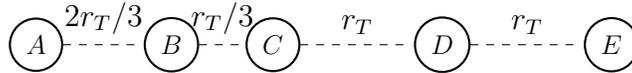
\begin{figure}[H]
\begin{center}
\begin{tikzpicture}[scale=0.9,shorten >=1pt, auto, node distance=1cm,
   node_style/.style={scale=0.85,circle,draw=black,thick},
   edge_style/.style={draw=black,dashed}]

    \node[node_style] (v1) at (0,0) {$A$};
    \node[node_style] (v2) at (2,0) {$B$};
    \node[node_style] (v3) at (3.5,0) {$C$};
    \node[node_style] (v4) at (6,0) {$D$};
    \node[node_style] (v5) at (8.5,0) {$E$};
    
    \draw[edge_style]  (v1) edge node{$2r_{T}/3$} (v2);
    \draw[edge_style]  (v2) edge node{$r_{T}/3$} (v3);
    \draw[edge_style]  (v3) edge node{$r_{T}$} (v4);
    \draw[edge_style]  (v4) edge node{$r_{T}$} (v5);
    
    \end{tikzpicture}
\end{center}
\vspace{-0.5cm}
\caption{A possible physical arrangement of a wireless network which leads to a claw-free conflict graph for Scenario I.}
\label{clf}
\end{figure}

\textbf{Example I:} A possible arrangement of wireless nodes to have a claw-free conflict graph is shown in Fig. \ref{clf}. Source and sink can be thought as the nodes $A$ and $E$, respectively. Let the maximum possible transmission distance be $r_{T}$ and $\Delta$ be very small. Then, we can model the conflict graph of this network as seen in Fig. 4.  The independent set polytope is the convex hull of the incidence vectors of the five independent sets $\{(A,B),(D,E)\}$, $(A,C)$, $\{(B,C),(D,E)\}$, $(C,D)$ and $(A,\{B,C\})$.

\begin{figure}[H]
\begin{center}
\begin{tikzpicture}[scale=2.4,shorten >=1pt, auto, node distance=1cm,
   node_style/.style={scale=0.45,circle,draw=black,thick},
   edge_style/.style={draw=black}]

    \node[node_style] (v1) at (0,0) {$(A,B)$};
    \node[node_style] (v2) at (1,0) {$(A,C)$};
    \node[node_style] (v3) at (0,-1) {$(A,\{B,C\})$};
    \node[node_style] (v4) at (1,-1) {$(B,C)$};
    \node[node_style] (v5) at (-0.5,-0.5) {$(C,D)$};
    \node[node_style] (v6) at (1.5,-0.5) {$(D,E)$};

    \draw[edge_style]  (v1) edge node{} (v2);
    \draw[edge_style]  (v1) edge node{} (v3);
    \draw[edge_style]  (v1) edge node{} (v4);
    \draw[edge_style]  (v1) edge node{} (v5);
    \draw[edge_style]  (v2) edge node{} (v3);
    \draw[edge_style]  (v2) edge node{} (v4);
    \draw[edge_style]  (v2) edge node{} (v5);
    \draw[edge_style]  (v2) edge node{} (v6);
    \draw[edge_style]  (v3) edge node{} (v4);
    \draw[edge_style]  (v3) edge node{} (v5);
    \draw[edge_style]  (v3) edge node{} (v6);
    \draw[edge_style]  (v4) edge node{} (v5);
    \draw[edge_style]  (v5) edge node{} (v6);

    \end{tikzpicture}
\end{center}
\vspace{-0.5cm}
\caption{Conflict graph of the network seen in Fig. \ref{clf}.}
\end{figure}
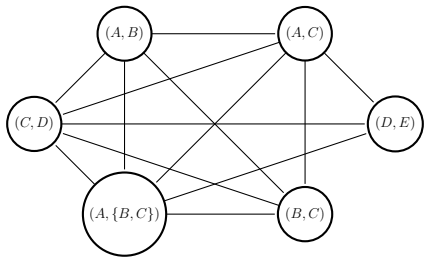

\vspace{-0.5cm}
We can generalize the Example I by realizing that the conflict graphs of the line networks are claw-free provided that there is enough distance between nodes to make 3 node away transmission impossible. We should physically satisfy $r_{T}<|P_{i}-P_{i+3}|$ for all $i\in \{1,2,...,n-3\}$ where $P_{i}$ is the position of the $i$th node in the network. This inequality can be easily satisfied by many different positioning scenarios, so, for simplicity, we can use a hypergraph $\mathcal{H}=(\mathcal{N},\mathcal{A})$ to represent our model, where $\mathcal{N}$ denotes the nodes and $\mathcal{A}$ denotes the hyperedges to symbolize valid transmissions between wireless nodes. Hypergraph of a line network, which has claw-free conflict graph, changes depending on the number of nodes to which a node $i$ can transmit, which can be $1$ or $2$ for every $i\in \{1,2,...,n-2\}$ and $1$ for $i\in \{n-1\}$. We have $2^{n-2}$ different possible hypergraphs of a line network, with $n$ nodes, all of which lead to claw-free conflict graphs. For instance, hypergraph of the network in Fig. 3 can be seen in Fig. 5. Here, first node is able to transmit up to next two neighbours whereas the other nodes are only able to transmit to next node. Transmissions $(A,B)$ and $(C,D)$ are not simultaneously possible, because we use the Protocol model to decide interference relations. Receiver $B$ is closer to $C$, a transmitter of another transmission, than to $A$ and this violates the constraint (\ref{eq:2}). Directed antennas, which we do not assume in our scenario, could be used to avoid the interference.

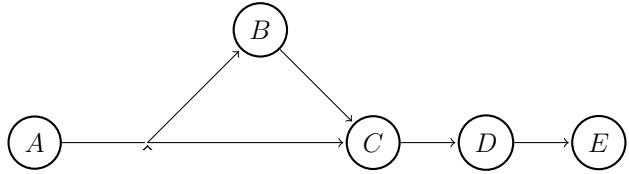
\begin{figure}[H]
\begin{center}
\begin{tikzpicture}[scale=1.5,shorten >=1pt, auto, node distance=1cm,
   node_style/.style={scale=0.85,circle,draw=black,thick},
   edge_style/.style={draw=black}]

    \node[node_style] (v1) at (0,0) {$A$};
    \node[node_style] (v2) at (2,1) {$B$};
    \node[node_style] (v3) at (3,0) {$C$};
    \node[node_style] (v4) at (4,0) {$D$};
    \node[node_style] (v5) at (5,0) {$E$};

    \draw[edge_style]  (v1) edge node{} (1,0);
    \draw[->]  (1,0) edge node{} (v2);
    \draw[->]  (1,0) edge node{} (v3);
    \draw[->]  (v2) edge node{} (v3);
    \draw[->]  (v3) edge node{} (v4);
    \draw[->]  (v4) edge node{} (v5);
    
    \end{tikzpicture}
\end{center}
\vspace{-0.5cm}
\caption{Hypergraph of the network seen in Fig. 3.}
\end{figure}

\textbf{Theorem 1:} \textit{Under the assumptions of Scenario I, the conflict graph of a line network where a transceiver is able to transmit to at most 2 nodes and all transceivers convey information in the direction from source to sink is guaranteed to be claw-free.}
\\

\begin{figure}[H]
\begin{center}
\begin{tikzpicture}[scale=1.5,shorten >=1pt, auto, node distance=1cm,
   node_style/.style={scale=0.7,circle,draw=black,thick},
   edge_style/.style={draw=black,dashed}]

    \node[node_style] (v1) at (0,0) {$A$};
    \node[node_style] (v2) at (1,0) {$B$};
    \node[node_style] (v3) at (2,0) {$C$};
    \node[node_style] (v4) at (3,0) {$D$};
    \node[node_style] (v5) at (4,0) {$E$};
    \node[node_style] (v6) at (5,0) {$F$};
    
    \draw[edge_style]  (v1) edge node{} (v2);
    \draw[edge_style]  (v2) edge node{} (v3);
    \draw[edge_style]  (v3) edge node{} (v4);
    \draw[edge_style]  (v4) edge node{} (v5);
    \draw[edge_style]  (v5) edge node{} (v6);
    \draw[->]  (v5) edge [bend right=30] (v6);
    \draw[->]  (v1) edge [bend right=30] (v2);
    \draw[->]  (v3) edge [bend right=30] (v4);
    
    \end{tikzpicture}
\end{center}
\vspace{-0.5cm}
\caption{Illustration for proof of Theorem 1.}
\end{figure}
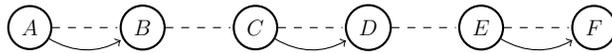

\vspace{-0.5cm}

\begin{proof}
Let us prove this theorem by contradiction. To have a claw in the conflict graph, a transmission $v_{1}$ should have conflicts with three other transmissions $v_{2}, v_{3}, v_{4}$ whereas those three should not have any conflicts between them. Let us use Fig. 6 for ease of understanding. Since transmissions are from source to sink, assume that a node only transmits to nodes that are located closer to sink in terms of hop distance. To this end, assume $v_{1}$ as the central node of a possible claw, $v_{1}=(C,D)$. Then, we can have one interfering transmission from the source side of $C$, say $v_{2}=(A,B)$, and one from the sink side of $D$, $v_{3}=(E,F)$ not to have interference between $v_{2}$ and $v_{3}$. $B$ and $E$ are chosen to be as far as possible. In such situation, transmitting nodes $C$ and $E$ cause interference on receiving nodes $B$ and $D$, respectively. Now, we have to place the transmitter and the receiver of the last transmission. This one has to interfere with $(C,D)$ without interfering with $(A,B)$ and $(E,F)$ to induce a claw in the conflict graph of the network. If we place the transmitter on the source side of $C$, this leads to an interference with $(A,B)$ which will break the claw, so this option is not possible. Also, we cannot place the receiver in the sink side of $D$ since this leads to an interference with $(E,F)$ which will again break the claw. Therefore, since the receiver must be on the sink side relative to the transmitter, the only remaining option is to place both the transmitter and the receiver between $C$ and $D$. However, this option makes $C$ able to transmit to 3 different nodes where we assume each node is able to transmit to at most 2 nodes.
\end{proof}

\subsection{Scenario II - Tree Networks}
We can also have other network topologies that lead to claw-free conflict graphs. One of them is a tree representation of the network, but since it is harder to get claw-freeness with the same assumptions for the line topology model, we propose new set of assumptions:

\begin{itemize}
\item A transceiver can receive from at most one transceiver in a time slot.
\item Time division duplex transceivers are used.
\item Nodes are arranged as a tree topology.
\item We directly work on hypergraph model without any consideration on physical locations of nodes and assume an interference model based on hops instead of the Protocol model.
\item Transmissions are in the direction from root node to leaves of tree.
\item Directed antennas are used, so interference can only occur in the forward direction along the tree.
\item A transmitting node does not lead to any interference to the receivers which are 3-hops or more away from it.
\item Only one node in every level can have children.
\end{itemize}

In the construction of conflict graph, let $v_{1}, v_{2}\in \mathcal{V}$ be in the conflict graph.$\{v_{1},v_{2}\}\in \mathcal{E}$ if any of the below conditions hold:

\begin{enumerate}[label=\textbf{C2.2.\arabic*},ref=C2.2.\arabic*,labelindent=20pt,itemindent=1em,leftmargin=25pt]
\item \label{c1} $i_{1}=i_{2}$.
\item \label{c2} $(i_{1}\in J_{2}) || (i_{2}\in J_{1})$.
\item \label{c3} $J_{1}\cap J_{2}\neq \emptyset$.
\item \label{c4} ($i_{1}$ is a child of $i_{2}$)||($i_{2}$ is a child of $i_{1}$). \label{last-item2}
\end{enumerate}

\textbf{Theorem 2:} \textit{Under the assumptions given in Scenario II, conflict graph of a wireless network is guaranteed to be claw-free.}

\begin{proof}
Let us assume that the root node belongs to level $1$ and a node, which has a distance $k$ to root in terms of hyperedge number, belongs to level $k+1$. Now, consider the transmission $v_{1}$, from a node $i_{k}$ in level $k$ to its children that reside in level $k+1$. We have $2^{N_{k}}-1$ different interfering transmissions to $v_{1}$ which are from level $k-1$ to $k$ where $N_{k}$ is the number of children of $i_{k}$'s parent. Since only one of these transmissions can be scheduled in one time slot, they induce a complete subgraph in the conflict graph. Therefore, we can only choose one transmission, say $v_{2}$, from level $k-1$ to $k$ which has interference with $v_{1}$ because the cardinality of maximum independent set of the complete graph is $1$. Assuming that we have the node $i_{k+1}$, which belongs to level $k+1$ and has children, in the receiver set of the transmission $v_{1}$ (otherwise, it is easier to say that we will not have a claw.), we have another complete subgraph which contains the transmissions from the node $i_{k+1}$ to its children in level $k+2$. One of these transmissions can be selected for a possible claw, say $v_{3}$. Since, it is not possible to find another independent transmission $v_{4}$, we have a claw-free conflict graph for the network.
\end{proof}

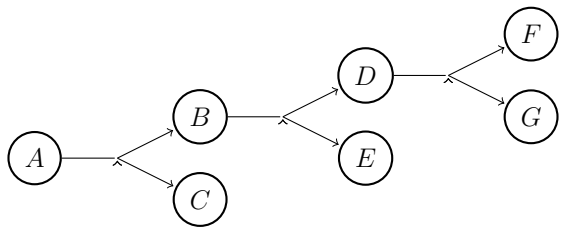
\begin{figure}[H]
\begin{center}
\begin{tikzpicture}[scale=1.1,shorten >=1pt, auto, node distance=1cm,
   node_style/.style={scale=0.85,circle,draw=black,thick},
   edge_style/.style={draw=black}]

    \node[node_style] (v1) at (0,0) {$A$};   
    \node[node_style] (v2) at (2,0.5) {$B$};
    \node[node_style] (v3) at (2,-0.5) {$C$};    
    \node[node_style] (v4) at (4,1) {$D$};
    \node[node_style] (v5) at (4,0) {$E$};
    \node[node_style] (v6) at (6,1.5) {$F$};  
    \node[node_style] (v7) at (6,0.5) {$G$};
    
    \draw[edge_style]  (v1) edge node{} (1,0);
    \draw[->]  (1,0) edge node{} (v2);
    \draw[->]  (1,0) edge node{} (v3);
    \draw[edge_style]  (v2) edge node{} (3,0.5);
    \draw[->]  (3,0.5) edge node{} (v4);
    \draw[->]  (3,0.5) edge node{} (v5);
    \draw[edge_style]  (v4) edge node{} (5,1);
    \draw[->]  (5,1) edge node{} (v6);
    \draw[->]  (5,1) edge node{} (v7);

    \end{tikzpicture}
\end{center}
\vspace{-0.5cm}
\caption{An example of a tree network which has a claw-free conflict graph for Scenario II.}\label{fiko1}
\end{figure}

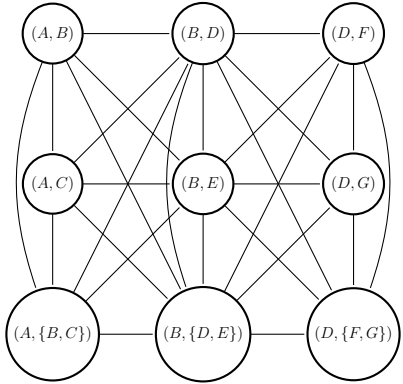
\begin{figure}[H]
\begin{center}
\begin{tikzpicture}[scale=2,shorten >=1pt, auto, node distance=1cm,
   node_style/.style={scale=0.5,circle,draw=black,thick},
   edge_style/.style={draw=black}]

    \node[node_style] (v1) at (-1,1) {$(A,B)$};
    \node[node_style] (v2) at (-1,0) {$(A,C)$};
    \node[node_style] (v3) at (-1,-1) {$(A,\{B,C\})$};
    \node[node_style] (v4) at (0,1) {$(B,D)$};
    \node[node_style] (v5) at (0,0) {$(B,E)$};
    \node[node_style] (v6) at (0,-1) {$(B,\{D,E\})$};
    \node[node_style] (v7) at (1,1) {$(D,F)$};
    \node[node_style] (v8) at (1,0) {$(D,G)$};
    \node[node_style] (v9) at (1,-1) {$(D,\{F,G\})$};
 
    \draw[edge_style]  (v1) edge node{} (v2);
    \draw[edge_style]  (v1) edge [bend right=20] (v3);
    \draw[edge_style]  (v2) edge node{} (v3);
    \draw[edge_style]  (v4) edge node{} (v5);
    \draw[edge_style]  (v4) edge [bend right=20] (v6);
    \draw[edge_style]  (v5) edge node{} (v6);
    \draw[edge_style]  (v7) edge node{} (v8);
    \draw[edge_style]  (v7) edge [bend left=20] (v9);
    \draw[edge_style]  (v8) edge node{} (v9);
    \draw[edge_style]  (v1) edge node{} (v4);
    \draw[edge_style]  (v1) edge node{} (v5);
    \draw[edge_style]  (v1) edge node{} (v6);
    \draw[edge_style]  (v2) edge node{} (v4);
    \draw[edge_style]  (v2) edge node{} (v5);
    \draw[edge_style]  (v2) edge node{} (v6);
    \draw[edge_style]  (v3) edge node{} (v4);
    \draw[edge_style]  (v3) edge node{} (v5);
    \draw[edge_style]  (v3) edge node{} (v6);
    \draw[edge_style]  (v4) edge node{} (v7);
    \draw[edge_style]  (v4) edge node{} (v8);
    \draw[edge_style]  (v4) edge node{} (v9);
    \draw[edge_style]  (v5) edge node{} (v7);
    \draw[edge_style]  (v5) edge node{} (v8);
    \draw[edge_style]  (v5) edge node{} (v9);
    \draw[edge_style]  (v6) edge node{} (v7);
    \draw[edge_style]  (v6) edge node{} (v8);
    \draw[edge_style]  (v6) edge node{} (v9);

    \end{tikzpicture}
\end{center}
\vspace{-0.5cm}
\caption{Conflict graph of the network seen in Fig. \ref{fiko1}.}\label{fiko2}
\end{figure}

An example of a tree network which has a claw-free conflict graph and its conflict graph can be seen in Fig. \ref{fiko1} and Fig. \ref{fiko2}, respectively.

\vspace{-1cm}

\vspace{-0.75cm}

Scenario II can be changed in order to relax the topology by letting every node have children as follows. Transmission and interference schemes are less restricted: We only allow for full-duplex transceivers and assume interference to nodes which are 2-hop away is not possible. Then, $\{v_{1},v_{2}\}\in \mathcal{E}$ if any of the below conditions hold:

\begin{enumerate}[label=\textbf{C2.3.\arabic*},ref=C2.3.\arabic*,labelindent=20pt,itemindent=1em,leftmargin=25pt]
\item \label{c1} $i_{1}=i_{2}$.
\item \label{c2} $J_{1}\cap J_{2}\neq \emptyset$.\label{last-item3}
\end{enumerate}

\begin{figure}[H]
\begin{center}
\begin{tikzpicture}[scale=0.8,shorten >=1pt, auto, node distance=1cm,
   node_style/.style={scale=0.75,circle,draw=black,thick},
   edge_style/.style={draw=black}]

    \node[node_style] (v1) at (0,0) {$A$};   
    \node[node_style] (v2) at (2,1) {$B$};
    \node[node_style] (v3) at (2,-1) {$C$};    
    \node[node_style] (v4) at (4,2) {$D$};
    \node[node_style] (v5) at (4,0) {$E$};
    \node[node_style] (v6) at (4,-2) {$F$};  
    \node[node_style] (v7) at (6,1) {$G$};
    \node[node_style] (v8) at (6,-1) {$H$};  
    \node[node_style] (v9) at (8,0) {$I$};
    
    \draw[edge_style]  (v1) edge node{} (1,0);
    \draw[->]  (1,0) edge node{} (v2);
    \draw[->]  (1,0) edge node{} (v3);
    \draw[edge_style]  (v2) edge node{} (3,1);
    \draw[->]  (3,1) edge node{} (v4);
    \draw[->]  (3,1) edge node{} (v5);
    \draw[edge_style]  (v3) edge node{} (3,-1);
    \draw[->]  (3,-1) edge node{} (v5);
    \draw[->]  (3,-1) edge node{} (v6);
    \draw[edge_style]  (v5) edge node{} (5,0);
    \draw[->]  (5,0) edge node{} (v7);
    \draw[->]  (5,0) edge node{} (v8);
    \draw[edge_style]  (v4) edge node{} (5,2);
    \draw[->]  (5,2) edge node{} (6,3);
    \draw[->]  (5,2) edge node{} (v7);
    \draw[edge_style]  (v6) edge node{} (5,-2);
    \draw[->]  (5,-2) edge node{} (v8);
    \draw[->]  (5,-2) edge node{} (6,-3);
    \draw[edge_style]  (v7) edge node{} (7,1);
    \draw[->]  (7,1) edge node{} (8,2);
    \draw[->]  (7,1) edge node{} (v9);
    \draw[edge_style]  (v8) edge node{} (7,-1);
    \draw[->]  (7,-1) edge node{} (v9);
    \draw[->]  (7,-1) edge node{} (8,-2);
    
    \node at ($(6,3)!.5!(7,3)$) {\ldots};
    \node at ($(8,2)!.5!(9,2)$) {\ldots};
    \node at ($(8,-2)!.5!(9,-2)$) {\ldots};
    \node at ($(6,-3)!.5!(7,-3)$) {\ldots};
    
    \end{tikzpicture}
\end{center}
\vspace{-0.5cm}
\caption{An example of a diamond network that leads to a claw-free conflict graph under the modified assumptions of Scenario II.}
\label{baklavaaa}
\end{figure}
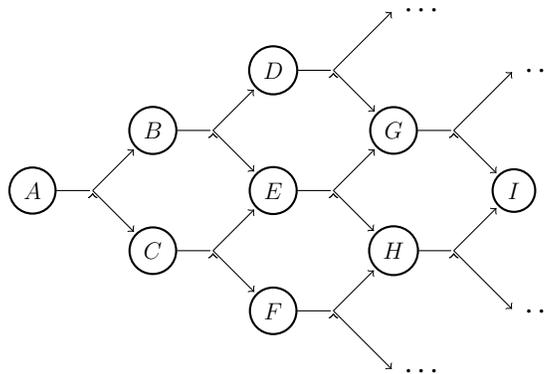

Within these assumptions, we can also introduce a family of networks called diamond networks which are claw-free except having tree topology. Diamond networks are scalable dense networks which exploit directed antennas. A typical diamond network can be seen at Fig. \ref{baklavaaa}. In a diamond network, every transceiver can transmit to at most two transceivers and can receive from at most two transceivers.

\vspace{-1cm}

\section{Methods to Reach Claw-freeness}
\label{sec:model2}
After some trials to see when we get a reasonable number of claws under different set of conditions, we conclude that we should do the following assumptions:

\begin{itemize}
\item We use Protocol model, therefore inequalities (\ref{eq:1}) and (\ref{eq:2}) are still valid.
\item A transceiver can listen to at most one transceiver at the same time.
\item Full duplex transceivers are used where we ignore self interference.
\item A transmission is possible if $x_{sender}< x_{receiver}$ in the 2-D coordinate system.
\item 60 degree directional antennas are deployed.
\end{itemize}

Under these assumptions, the conditions for building the conflict graph change. Say $S_{1},S_{2}\in \mathcal{V}$, then $\{S_{1},S_{2}\}\in \mathcal{E}$ if any of the following conditions hold for $S_{1} = (i_1,J_1)$ and $S_{2} = (i_2,J_2)$, which means they cannot be scheduled for the same time slot:

\begin{enumerate}[label=\textbf{C5.\arabic*},ref=C5.\arabic*,labelindent=20pt,itemindent=1em,leftmargin=18pt]
\item \label{2c1} $i_{1}=i_{2}$.
\item $J_{1}\cap J_{2}\neq \emptyset$.
\item $|i_{2}-j|\leq (1+\Delta)|i_{1}-j|$ $\&$ $\Bigl|\arctan\Bigl(\dfrac{y_j-y_{i_2}}{x_j-x_{i_2}}\Bigr)\Bigr| < \dfrac{\pi}{6}  $ for $\exists j\in J_{1}$.
\item $|i_{1}-j|\leq (1+\Delta)|i_{2}-j|$ $\&$ $\Bigl|\arctan\Bigl(\dfrac{y_j-y_{i_1}}{x_j-x_{i_1}}\Bigr)\Bigr| < \dfrac{\pi}{6}  $ for $\exists j\in J_{2}$.
\end{enumerate}

\begin{table*}[t]
\centering
    \begin{tabular}{ | p{3.2cm} | p{3.2cm} | p{3.2cm} | p{3.2cm} |p{3.2cm}|}
    \hline
    Transmission range & Average number of claws in the conflict graph & Connectedness of the network & Claw-freeness of the conflict graph & Average number of transmissions in the network \\ \hline
    $r_T=7$ & 0.19 & Unconnected & Claw-free in 96\% of trials & 5.47\\ \hline
    $r_T=8$ & 0.99 & Connected in 1\% of trials & Claw-free in 95\% of trials & 6.89\\ \hline
    $r_T=9$ & 3.13 & Connected in 1\% of trials & Claw-free in 84\% of trials & 11.43\\ \hline
    $r_T=10$ & 15.61 & Connected in 4\% of trials & Claw-free in 83\% of trials & 11.38\\ \hline
    $r_T=11$ & 24.44 & Connected in 4\% of trials & Claw-free in 68\% of trials & 15.69\\ \hline
    $r_T=12$ & 15.75 & Connected in 10\% of trials & Claw-free in 68\% of trials & 18.49\\ \hline
    $r_T=13$ & 23.30 & Connected in 17\% of trials & Claw-free in 56\% of trials & 23.23\\ \hline
    $r_T=14$ & 39.43 & Connected in 19\% of trials & Claw-free in 53\% of trials & 25.39\\ \hline
    \end{tabular}
    
\caption{Simulation results for randomly generated transceiver locations ($n=10$). Ratio of intersection between connectedness and claw-freeness is $r_T=10\rightarrow 0.01$, $r_T=11\rightarrow 0.01$, $r_T=12\rightarrow 0.03$, $r_T=13\rightarrow 0.05$, $r_T=14\rightarrow 0.07$.} \label{teybil2}
\vspace{-1cm}
\end{table*}

To assess the introduction of claws in the conflict graph, we set up a 2-D coordinate system and randomly assign coordinates to $n$ transceivers in an area of $n^{2}$ where we set $n=10$. Simulations are done 100 times for each $r_T$ value and averaged in the end. Results can be seen in Table \ref{teybil2}. We observe that we almost always get an unconnected network in cases that we have very low number of claws in average as seen for $r_T=7,8,9$. According to varying transmission range $r_T$, we observe a rapid increase in the number of claws when connected networks start to appear for $r_T=10$. Note that, intersection ratio of connectedness and claw-freeness increases when $r_T$ increases although claw-freeness ratio decreases. On the other hand, average number of claws appearing in a conflict graph increases with $r_T$. Therefore, there is obviously an important connectedness claw-freeness trade-off. We observe that in a scenario where transceivers have a low transmission range $r_T$, we usually get a claw-free conflict graph in the expense of connectedness of given network. In case of a relatively higher transmission range $r_T$, we can easily get a connected network, but also having plenty of claws in the conflict graph. In such a case, it may seem impractical to get rid of high number of claws without having much effect on MWIS. However, even one action like a very small position change or an introduction of a conflict edge to conflict graph can break tens of claws and leads to satisfying results as it can be seen in Section \ref{simulasyon}.

\begin{figure}[H]
  \centering
    \includegraphics[width=0.4\textwidth]{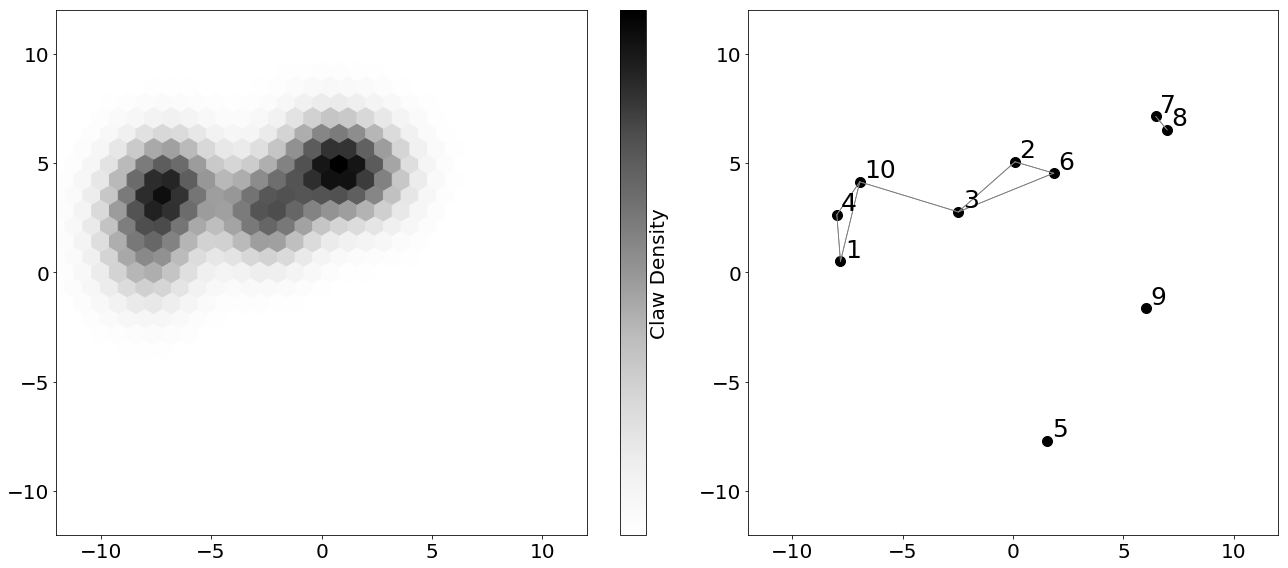}    
    \caption{Left figure shows claw density represented with a heat map. Right figure shows the network which consists of transceivers and their connections.}
    \label{heat_map}
\end{figure}

An example network where we have $n=10$ transceivers can be seen in Fig. \ref{heat_map}. In this illustration, we deploy omnidirectional antennas with time division duplex transceivers. In the right part of Fig. \ref{heat_map}, the locations of transceivers are shown in a 2-D coordinate system. An edge between two transceivers $i_a$ and $i_b$ means that $|P_{i_a}-P_{i_b}|\leq r_{T}$ where $P_{i_a}$ denotes the position of transceiver $i_a$ in the coordinate system. As seen, we do not have a path between every pair of transceivers, so we have an unconnected network in this example. A claw has 4 different vertices which can be represented with $(i_t,J_t)$. In the left figure, claw density of the network can be seen where we assign equal weights to transmitters and receivers in the transmission vertices creating the claw. To be more clear, we assign same weights to transmitters $i_1,i_2,i_3,i_4$ and to receivers which are included in sets $J_1,J_2,J_3,J_4$ where $(i_t,J_t)$ defines a transmission which belongs to the claw for $t$ $\in$ $\{1,2,3,4\}$.

\vspace{-0.75cm}

\begin{figure}[H]
  \centering
    \includegraphics[width=0.4\textwidth]{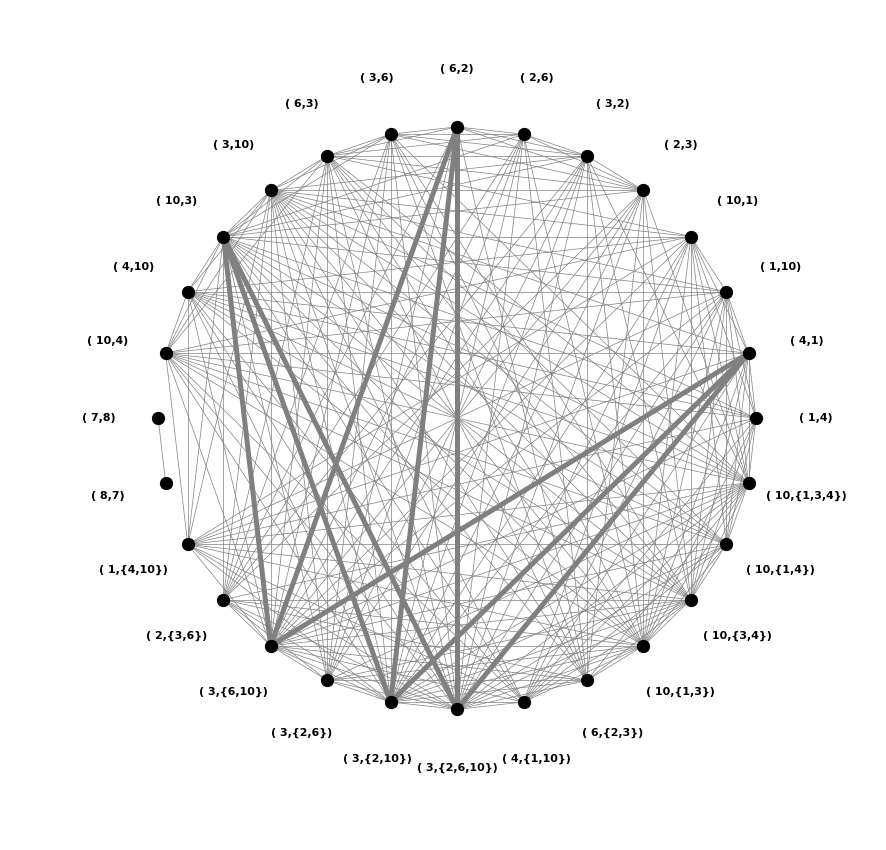} 
    \vspace{-0.5cm}
    \caption{Conflict graph of the network given in Fig. \ref{heat_map}. Vertices represent transmissions and edges represent conflicts. Bold edges represent the claws in the conflict graph.}
\label{fig:guici}
\end{figure}

In Fig. \ref{fig:guici}, we give the conflict graph of the network seen in Fig. \ref{heat_map}. In this conflict graph, we have 28 possible transmissions and 3 different induced claws. We assume that if there are $M$ pairwise independent vertices connected to a central vertex, there are ${M}\choose{3}$ different claws in this induced subgraph. Induced claws can be seen at Fig. \ref{fig:guici} with bold lines. 

\vspace{-0.5cm}

\vspace{-0.75cm}

Our goal is to achieve claw-freeness in conflict graph by doing necessary modifications under the assumptions of Section \ref{sec:model2}. To this end, using transceivers' coordinates and transmission range of devices, we model the conflict graph of a given network. After modeling the conflict graph, we test the claw-freeness and if the conflict graph is not claw-free, we find all the claws in the conflict graph and spot the locations of transceivers that play role in the resulting claws. Then, we aim to make a given ad hoc network suitable for polynomial time scheduling. To this end, network's conflict graph must satisfy claw-freeness property. What can we do to achieve our goal? We have two different approaches:

\begin{itemize}
\item Directly breaking claws on conflict graph.
\item Making physical modifications in network.
\end{itemize}

\section{Method I - Breaking claws on Conflict Graph}
\label{breakclaw}

Given a wireless network, we know how to model its conflict graph using possible transmissions and interference between them. Let us assume that the resultant conflict graph contains claws. Also, if there are strict constraints on the physical alignment of the network originating from the nature of the application and we are not allowed to make modifications, then we must find another solution to pave the way for claw-freeness in the conflict graph. Here, we propose an approach that modifies the conflict graph, $\mathcal{G}=(\mathcal{V}, \mathcal{E})$, without making any changes in the network configuration. What we do is to add necessary edges to conflict graph to make it claw-free. In other words, we pretend that some pair of transmissions, say $S_x = (i_x,J_x)$ and $S_y = (i_y,J_y)$, cannot be scheduled for the same time slot even if they do not interfere. So, we modify the conflict graph such that $\{S_{x},S_{y}\}\in \mathcal{E}$ where $S_{x},S_{y}\in \mathcal{V}$.

\iffalse
    This solution induces another question. Between which nodes of the conflict graph should we introduce edges? Fortunately, we can automatize this process, our heuristic solution is given below:
    
    \begin{itemize}
    \item Find all claws in the conflict graph $\mathcal{G}=(\mathcal{V}, \mathcal{E})$. If there is not any, terminate the algorithm. Else, go to next step.
    \item Since a claw is made of 4 nodes, 1 center and 3 surrounding nodes, calculate a vector $\vec{C}$ of nodes where $\vec{C}$ is given according to non-increasing number of occurrences of nodes in surrounding parts of claws.
    \item Find $(i,j)$ pair satisfying $\{\vec{C}[i],\vec{C}[j]\}\notin \mathcal{E}$ and minimizing $i+j$.
    \item If there are multiple $(i,j)$ pairs satisfying the step above, choose one which minimizes $abs(i-j)$.
    \item Introduce the edge $\{\vec{C}[i],\vec{C}[j]\}\in \mathcal{E}$
    \item Go back to first step.
    \end{itemize}
\fi

\textbf{Example of introducing edge:} According to the conflict graph at Fig. \ref{fig:guici}, 3 claws can be broken by introducing only one edge between transmission nodes $(10,3)$ and $(6,2)$ although they normally can be done simultaneously without interference. However, note that adding an edge between two nodes might introduce new claws. These claws can be broken iteratively until the conflict graph is claw-free. In this case, breaking initial claws does not introduce new claws and conflict graph becomes claw-free. In other words, transmissions $(10,3)$ and $(6,2)$ cannot be scheduled for the same time slot if we want to achieve polynomial time scheduling. 

\begin{comment}

\textbf{Example II of introducing edge:} Suppose we have the network seen in Fig. \ref{fig:guici31}. This network does not satisfy the assumption of directional antennas with $\pi/3$ beam width which is given in C5.3 and C5.4, but for the sake of a simple example, we can ignore this fact for this example. As seen in Fig. \ref{fig:guici69}, its conflict graph contains 2 claws centered at $(1,\{3,4\})$ and $(1,\{2,3,4\})$. We can simultaneously break these claws by introducing an edge between arbitrarily chosen two elements of the set $\{(1,2),(2,3),(3,4)\}$. Again, we observe that it is possible to get a claw-free conflict graph by introducing an edge on it and without any interventions to the network setup.

\begin{figure}[H]
  \centering
    \includegraphics[width=0.35\textwidth]{yaralistyla.png}    
    \caption{An example network for illustration of claw breaking}
\label{fig:guici31}
\end{figure}

\begin{figure}[H]

  \centering
    \includegraphics[width=0.43\textwidth]{claw_break4.png}    
    \caption{Conflict graph of the network seen in Fig. \ref{fig:guici31}. Bold edges represent the claws.}
\label{fig:guici69}
\end{figure}

\end{comment}

Beyond such simple examples, we shall investigate how to automate the process of deciding where to introduce new edges so that the decrease in the weighted size of MWIS is minimal.

%METHODOLOGY
\iffalse
As we have explored, the problem of finding maximum weighted independent sets (MWIS) has exponential-time complexity in general. However, in claw-free graphs, MWIS can be identified in polynomial-time. For this reason, we propose to convert our conflict graph into a claw-free graph by introducing additional conflicts, i.e. edges in the graph. As these newly-added conflicts do not actually exist in our system, the MWIS we identify will be sub-optimal in terms of its weighted size. Therefore, our algorithm needs to eliminate the claws in the graph with minimal decrease in the MWIS weighted size. Now, we present the algorithm in detail.
\fi

Given a conflict graph $G = (V,E)$, we denote the set of all missing edges as $\tilde{E}$. We borrow the intuition behind steepest descent optimization
%[4]
and construct the following greedy algorithm. Starting with $G$, at each iteration, we identify the edge $e \in \tilde{E}$ such that the action $E \leftarrow E \cup \{e\}$ causes a decrease in the quantity of G's claws as much as possible per decrease in the G's MWIS weighted size. If there exists more than one such edges $e$ we sample amongst them uniformly and conclude with $E \leftarrow E \cup \{e\}$, $\tilde{E} \leftarrow \tilde{E} \backslash \{e\}$. We continue until all the claws in $G$ is eliminated.

\subsection{Substitute for MWIS weighted size}
The proposed algorithm requires to calculate how the addition of an edge decreases the MWIS weighted size. However, to our knowledge, the only way is to identify the MWIS both before and after the addition of the said edge. Since the motivation of the algorithm was to avoid the actual calculation of the MWIS in non-claw-free graphs, we need to utilize a meaningful substitute.

Our aim is to eliminate the number of claws with as few edge introductions as possible. Hence, we assume that the structure of our graph changes minimally implying the histogram of maximal independent set weighted sizes maintains its initial form. Through this thinking process, we propose to substitute expected maximal independent set weighted size, which we denote as $S_I$, for the actual MWIS weighted size.

\begin{algorithm2e}
    \small
	\DontPrintSemicolon
	\SetAlgoLined
	\caption{Generation of a Maximal Independent Set by processing an Ordered Vertex Set}\label{algomaximal}
	\SetKwInOut{Input}{Input}
	\SetKwInOut{Output}{Output}
	%\SetKwInOut{Initialize}{Initialize}
	\Input{$U = \{u_1, ..., u_N\}$}
	\Output{A Maximal Independent Set $S_U$}
	\BlankLine
	%\Initialize{$S_U=\emptyset$}
	Initialize $S_U=\emptyset$.\\
	\For{$i=1,...,N$}{
		\If{$u_i$ is not connected to any $s \in S_U$}{
			$S_U \leftarrow S_U \cup \{u_i\}$	
		}
	}
	Output $S_U$.
\end{algorithm2e}

First, we identify a given ordering of the vertex set $V$ as $U = \{ u_1, ..., u_N \}$ where $N = |V|$. Then, Algorithm \ref{algomaximal} outputs a maximal set $S_U$ heavily dependant on the ordering provided by $U$. Finally, we calculate $S_I = \mathbf{E}_U[|S_U|_w]$. Here, the expectation is uniformly over all possible $U$'s for a given vertex set $V$ and the set size operation $|\cdot|_w$ accounts for the vertex weights. This makes $S_I$ the expected maximal independent set weighted size.

Another way to write $S_I$ is such that,
\begin{align}
	S_I &= \mathbf{E}_U \left[ |S_U|_w \right] \nonumber
	\\&= \mathbf{E}_U \left[\sum_{v\in V} w(v) \cdot \mathbbm{1}_{v \in S_U} \right] \nonumber
	\\&= \sum_{v\in V} w(v) \cdot \mathbf{E}_U \left[ \mathbbm{1}_{v \in S_U} \right], \label{S_I}
\end{align}
where $w: V \rightarrow \mathbb{R}$ is the weight mapping for the vertices and $\mathbbm{1}_{v \in S_U}$ is the indicator function for the random event of vertex $v$ being included in $S_U$, i.e. $\mathbbm{1}_{v \in S_U} = 1$ if $v\in S_U$ and $\mathbbm{1}_{v \in S_U} = 0$ otherwise. The randomness originates from choosing a random permutation $U$ of the set $V$ to generate $S_U$.

The expectation of an indicator function is the probability of occurrence for the corresponding event. Thus,
\begin{equation} \label{v_in}
	\mathbf{E}_U \left[ \mathbbm{1}_{v \in S_U} \right] = \mathbf{P} \left( v \in S_U \right).
\end{equation}

Continuing from Eq. \ref{S_I} and \ref{v_in},
\begin{equation} \label{S_I_2}
	S_I 
	= \sum_{v\in V} w(v) \cdot \mathbf{P} \left( v \in S_U \right)
\end{equation}

To solve for $\mathbf{P}(v \in S_U)$, we define the neighbourhood of $v$ as $N_v$ such that each $u \in N_v$ shares an edge with $v$ (no loops, i.e. v does not share an edge with itself). We also define the degree of vertex $v$ as $d_v = |N_v|$.

Let us consider the indexing function $i_u: V \rightarrow \{1,...,|V|\}$ for an ordered vertex set $U$. $i_u(\cdot)$ is such that $v = u_{i_u(v)} \in U$. Given the set $U$, the vertex $v$ is sure to be chosen for $S_U$ if $i_u(v) < i_u(v')$, for every $v' \in N_v$. For the other placements of $v$ in $U$, the effect on $S_I$ is not trivial to decompose in terms of the vertex weights and degrees. Consequently, we neglect such additional components contributing to $S_I$.

Since we have considered all possible $U$'s as equally probable, the probability of $i_u(v) < i_u(v')$ for every $v' \in N_v$ is $1/(d_v+1)$. Consequently, using Eq. \ref{S_I_2},
\begin{equation} \label{s_lb}
	S_I 
	\geq \sum_{v\in V} \frac{w(v)}{d_v + 1}.
\end{equation}

In Eq. \ref{s_lb}, the lower-bound on RHS is actually a vertex weighted version of the Caro-Wei bound which is, to our knowledge, the unique quantity amongst both lower and upper bounds for the maximum independent set size easily decomposable into individual contributions from the vertices. Consequently, we conclude this part by setting the following expected maximum independent set weighted size contributions $S_v$,
\begin{equation}
	S_v = \frac{w(v)}{d_v + 1}
\end{equation}

\subsection{Claw-Freeing Algorithm}
Returning to our claw-freeing algorithm, for an edge $e \in \tilde{E}$, we identify its endpoints $v$ and $v'$, i.e. $e = (v,v')$. Let us denote the expected maximal independent set weighted size contributions after the operation $E \leftarrow E \cup \{e\}$, $\tilde{E} \leftarrow \tilde{E} \backslash \{e\}$ as $S_v^+$ for each $v \in V$. Let us denote the total change from $\sum_{v\in V} S_v$ to $\sum_{v\in V} S_v^+$ as $M_e$, such that,
\begin{align}
	M_e 
	&= \sum_{v\in V} S_v - \sum_{v\in V} S_v^+ \nonumber
	\\&= \left( S_v - S_v^+ \right) + \left( S_{v'} - S_{v'}^+ \right)
	\\&= \frac{w(v)}{d_v+1} - \frac{w(v)}{d_v+2} + \frac{w(v')}{d_{v'}+1} - \frac{w(v')}{d_{v'}+2} \nonumber
	\\&= \frac{w(v)}{(d_v+1)(d_v+2)} + \frac{w(v')}{(d_{v'}+1)(d_{v'}+2)} \label{M_e}
\end{align}

Let us denote the claw counting function as $C(\cdot)$ where for a graph $G$, $C(G)$ is the number of distinct claws present in the said graph.

As a reminder, given 4 vertices from the vertex set $V$ of the graph $G$, we claim they induce a claw
\iffalse
under the following condition. After labelling the said vertices as $v_1,v_2,v_3,v_4$ in a decreasing order according to their degrees, the edge set $E$ of the graph $G$ should only include the edges $(v_1,v_2), (v_1,v_3), (v_1,v_4)$ and not any other edges of the form $(v_i,v_j)$ where $1 \leq i \leq j \leq 4$. We denote such a complete bipartite graph as $K_{1,3}$.
\fi
only if their induced sub-graph is a $K_{1,3}$ complete bipartite graph.

Now, we define the quantity $\Delta_e$ as the decrease in the quantity of claws after the addition of $\{e\}$ into the graph $G=(V,E)$, i.e., 
\begin{equation}
	\Delta_e = C((V,E)) - C((V,E\cup\{e\})).
\end{equation}

Consequently, since we would like to add the missing edge $e \in \tilde{E}$ maximizing the decrease in claw per reduction in the expected maximal independent set weighted size, we choose $e$ such that,
\begin{equation}
	e = \argmax_{e' \in \tilde{E}} \frac{\Delta_{e'}}{M_{e'}}.
\end{equation}

Assuming strictly positive vertex weights, for the special case when there is no $e \in \tilde{E}$ such that $\Delta_e > 0$, then it means whichever edge we choose to add, our claw count will increase. In that case, we choose edge $e$ which eliminates the highest number of claws currently on graph $G = (V,E)$ even though it introduces more claws than it erases. This strategy is analogous to local optima escape tactics employed in the iterative optimization problems.

\subsection{Computational Cost Analysis of the Algorithm}
Throughout the claw-freeing algorithm, the quantities we need to keep track of are $d_v$, $S_v$, $M_e$ and $\Delta_e$ for each missing edge $e \in \tilde{E}$.

We calculate $d_v$ for every $v \in V$ in $O(\sum_{v\in V} (1+d_v))$ computational time. In terms of the total number of nodes $N=|V|$, this becomes $O(N^2)$. After each introduction of an edge $e = (v_1,v_2)$, we only increment $d_{v_1}$ and $d_{v_2}$. Since we can at most introduce $O(N^2)$ new edges, the total computational time spent on $d_v$ calculation is still $O(N^2)$.

Calculation of $S_v$ for every $v \in V$ requires $O(N)$ time after calculating every $d_v$. Like $d_v$, the computation time per edge introduction is constant for updating $S_v$, thus the total computational time spent on $S_v$ is also $O(N^2)$.

After calculating $d_v$ and $S_v$, the calculation of $M_e$ for every missing $e \in \tilde{E}$ is $O(N^2)$. However, unlike before, updating $M_e$ quantities has computational cost $O(N-d_{v_1} + N-d_{v_2})$ after the introduction of $e=(v_1,v_2)$ since we need to update $M_e$ for every $e \in \tilde{E}$ and the number of neighbours vertex $v$ has in the complement graph $\tilde{G} = (V,\tilde{E})$ is $(N-d_v)$. As a result, the overall computational cost of calculating $M_e$ is $O(N^3)$.

Calculation of the quantities $\Delta_e$ for each missing edge $e \in \tilde{E}$ requires identifying each unique claw in the graph. Hence, the computational cost is $O(\#Claws)$. As we have observed, the claw, i.e. $K_{1,3}$, is a three-pronged structure where we have a ternary tree of $4$ nodes with $1$ parent (root) and $3$ children. For every node $v$, the identification of a $K_{1,3}$ can be done in $O({d_v}^3)$ time. Consequently, the overall time is $O(\sum_{v\in V} {d_v}^3)$.

Furthermore, the calculation of $\Delta_e$ also requires identifying unique instances of another structure we shall call a "pre-claw". It occurs when out of $4$ vertices, $3$ form the two-pronged version of a claw, $K_{1,2}$, while the other is disconnected from the first three. Computational cost of finding the pre-claws is $O(\sum_{v\in V} {d_v}^2 |V|)$. The difference of changing one $d_v$ multiplier with $N$ compared to distinct claw identification results from the fact that after identifying a $K_{1,2}$, we also need to find a fourth vertex disconnected from the first three.

In terms of the total number of nodes $N$, the overall computational time -claw and pre-claw identifications combined- is at worst $O(N^4)$. In the worst-case, we may add $O(N^2)$ edges to eliminate the claws resulting in the overall computational time of $O(N^6)$ for obtaining a claw-free graph.

The computational cost attributed to $d_v$, $S_v$ and $M_e$ together is $O(N^3)$. However, the cost attributed to $\Delta_e$ is $O(N^6)$. Since the initial calculation cost of $\Delta_e$ was also larger than the cost of $d_v$, $S_v$ and $M_e$ with $O(N^4)$, we conclude that overall computational cost of the algorithm is dominated by the cost of computing $\Delta_e$ and currently is $O(N^6)$.

Although polynomial, this computational time is higher than the cost of state-of-the-art algorithms for finding MWIS in claw-free graphs.
%[Ref]
The pseudo-code for this part of the initial calculations can be found in Algorithm \ref{claw_full_algo}.

\begin{algorithm2e}
    \small
	\DontPrintSemicolon
	\caption{Claw-Freeing Algorithm - Initialization}\label{claw_full_algo}
	\SetKwInOut{Input}{Input}
	\SetKwInOut{Output}{Output}
	\Input{Initial Conflict Graph $G = (V,E)$ \\ Weight Mapping $w: V \rightarrow \mathbb{R}$}
	\Output{Claw-Free Graph $G_c$}
	\BlankLine
	Set $N = |V|$.\\
	
	Initialize $\Delta_e = 0$ for every $e \in \tilde{E}$.\\
	
	Initialize the number of claws $C = 0$.\\
	
	Initialize claw elimination counts $\Delta_e^* = 0$.
	
	NOTE: $(u,v)$ and $(v,u)$ are the same undirected edge.\\
	\ForEach{$v \in V$}{
		Identify neighbor set $N_v$.\\
		Calculate vertex degree $d_v = |N_v|$.\\
		Calculate size contribution $S_v = w(v)/(d_v+1)$.\\
		\ForEach{$\{u_1,u_2,u_3\} \subset N_v$}{
			Set $E_c = \{(u_1,u_2),(u_1,u_3),(u_2,u_3)\}$.\\
			\If{$\forall e \in E_c$, $e \notin E$ (i.e., a claw)}{
				\ForEach{$e \in E_c$}{
					$\Delta_e \leftarrow \Delta_e + 1$.\\
					$\Delta_e^* \leftarrow \Delta_e^* + 1$.
				}
				$C \leftarrow C + 1$.
			}
		}
		\ForEach{$\{u_1,u_2\} \subset N_v$}{
			\If{$(u_1,u_2) \notin E$}{
				\ForEach{$v_2 \in V \backslash (N_v \cup N_{u_1} \cup N_{u_2})$}{
					$\Delta_e \leftarrow \Delta_e - 1$, for $e=(v,v_2)$.
				}
			}
		}
	}
	\ForEach{$\{v,v'\} \subset V$ s.t. $(v,v') \notin E$}{
		$e = (v,v')$.\\
		$M_e = S_v/(d_v+2) + S_{v'}/(d_{v'}+2)$.\\
	}
\end{algorithm2e}

Our claw-freeing algorithm introduces a substantial bottleneck if $\Delta_e$ is calculated from scratch after each edge introduction. Therefore, we propose the following approach to efficiently calculate $\Delta_e$.

\subsection{Iterative Calculation of $\Delta_e$}
The introduction of a new edge $e \in \tilde{E}$, eliminates existing claws and introduces new claws only if the two out of four vertices involved in the eliminated or introduced claws are the endpoints of our new edge $e$. Therefore, it should be possible to reduce the per new edge computational cost of updating $M_e$ to $O(N^2)$, effectively resulting in total update cost of $O(N^4)$. Thus, even including the initial calculation cost of $O(N^4)$, overall computational cost of becomes $O(N^4)$. We will now detail how this can be achieved.

At each iteration of the edge introduction algorithm, we need to identify unique instances conforming to one of the following five types. Consider the newly introduced edge as $e=(v_1,v_2)$. After the identification of these instances, we follow with the provided updating of $\Delta_e$.

\vspace*{0.2cm}
\subsubsection{Type 1 - Claw Before}\hspace*{\fill}

Before the introduction of edge $e$, we had a claw such that $v_1$ and $v_2$ were two of the three children. This means, we had a root $u_1$ which was a neighbour of both $v_1$ and $v_2$. Furthermore, we had another child $u_2$ which was neighbour of $u_1$ but not $v_1$ and $v_2$

After introducing $e=(v_1,v_2) \in \tilde{E}$, the claw is broken. Hence, $\Delta_{e'}$ for the missing edges $e'=(v_1,u_2)$ and $e'=(v_2,u_2)$ are both decremented by 1.

\vspace*{0.2cm}
\subsubsection{Type 2 - Claw After}\hspace*{\fill}

Before the edge $e$, we had a pre-claw $K_{1,2}$ such that $u_1$ and $u_2$ are the children of $K_{1,2}$ while either $v_1$ or $v_2$ is the parent. This means $u_1$ and $u_2$ are disconnected and only one of the endpoints are neighbours with both while the remaining endpoint is disconnected from the other three.

After the edge $e=(v_1,v_2)$, a new claw is formed. We determine which of the endpoints ($v_1$ or $v_2$) is the root (parent) and denote the other as $v_*$. Afterwards, $\Delta_{e'}$ for $e'=(u_1,v_*)$, $e'=(u_2,v_*)$ and $e'=(u_1,u_2)$ are incremented by 1. 

\vspace*{0.2cm}
\subsubsection{Type 3 - Pre-Claw Before with Children Endpoints}\hspace*{\fill}

Before the edge $e$, we had a pre-claw $K_{1,2}$ where $v_1$ and $v_2$ are the children while $u_1$ is the parent. Note, $u_1 \in N_{v_1}$ and $u_1 \in N_{v_2}$. We also had $u_2$ which is disconnected with the other three.

After the edge $e$, the pre-claw is no more as $v_1$, $v_2$ and $u_1$ forms a triangle. Therefore, $e'=(u_1,u_2) \in \tilde{E}$ no longer introduces a new claw. Hence, $\Delta_{e'}$ is incremented by 1 to neutralize a previous reduction for the claw introduction for when $e'$ were to be introduced.

\vspace*{0.2cm}
\subsubsection{Type 4 - Pre-Claw Before with a Non-Child Endpoint}\hspace*{\fill}

Before the edge $e$, we had a pre-claw $K_{1,2}$ where one of the children is $v_1$ or $v_2$ and the parent is $u_1$. The remaining endpoint, temporarily denoted by $v_*$, is disconnected from all the previous vertices involved in $K_{1,2}$.

After the edge $e$, the pre-claw is no more as the $4$ vertices form a path now. Thus, $e'=(u_1,v_*) \in \tilde{E}$ no longer introduces a new claw and $\Delta_{e'}$ is incremented by 1 like in Type 3. 

\vspace*{0.2cm}
\subsubsection{Type 5 - Pre-Claw After}\hspace*{\fill}

Before the edge $e$, we have a structure consisting of $4$ vertices $\{v_1,v_2,u_1,u_2\}$ and $1$ edge which connects one of the endpoints, temporarily denoted as $v_*$, with one of its neighbours. $u_2$ is disconnected from the other three, similarly for the remaining endpoint. 

After the edge $e$, we obtain a pre-claw with $v_*$ at the root. Thus, $\Delta_{e'}$ for $e'=(v_*,u_2)$ is decremented by 1.

\vspace*{0.1cm}

To sum up, the iterative calculation of $\Delta_e$ decreases the computational cost calculating $\Delta_e$ from $O(N^6)$ to $O(N^4)$. This, in turn, reduces the overall computational cost of the claw-freeing algorithm to $O(N^4)$ since $\Delta_e$ calculation cost still dominates. Pseudo-code for the general run-time is displayed in Algorithm \ref{claw_full_algo_cont}.

\begin{algorithm2e}
    \small
	\DontPrintSemicolon
	\setcounter{AlgoLine}{23}
	\caption{Claw-Freeing Algorithm - Run-time}\label{claw_full_algo_cont}
	\SetKwRepeat{Do}{do}{while}
	\While{$C > 0$}{
		$e = \argmax_{e' \in \tilde{E}} \Delta_{e'}/M_{e'}$.\\
		\If{$\Delta_e \leq 0$}{
			$e = \argmax_{e' \in \tilde{E}} \Delta_{e'}^*$.
		}
		$E \leftarrow E \cup \{e\}$. \tab $C \leftarrow C - \Delta_e$.\\
		Identify endpoints $(v,v') = e$.\\
		\ForEach{$v^* \in \{v,v'\}$}{
		    $d_{v^*} \leftarrow d_{v^*} + 1$.\\
		    $S_{v^*} \leftarrow S_{v^*} \cdot d_{v^*}/(d_{v^*}+1)$.\\
		    \ForEach{$u \in V \backslash \{v^*\}$ s.t. $(v^*,u) \notin E$}{
			    $M_{e'} = S_{v^*}/(d_{v^*}+2) + S_u/(d_u+2)$, for $e' = (v^*,u)$.\\
		    }
		}
		\ForEach{$u' \in V \backslash (N_v \cup N_{v'})$}{
			\ForEach{$u \in N_v \cap N_{v'}$}{
				\eIf{$(u,u') \in E$}{
					$\Delta_e \leftarrow \Delta_e - 1$ for $e \in \{(v,u'),(v',u')\}$.\\
					$\Delta_e^* \leftarrow \Delta_e^* - 1$ for $e \in \{(v,u'),(v',u')\}$.
				}{
					$\Delta_e \leftarrow \Delta_e + 1$ for $e = (u,u')$.
				}
			}
			\ForEach{$\{v^*,v^-\} \in \{\{v,v'\},\{v',v\}\}$}{
    			\ForEach{$u \in N_{v^*} \backslash N_{v^-}$}{
    				\eIf{$(u,u') \in E$}{
    					$\Delta_e \leftarrow \Delta_e + 1$ for $e = (v^-,u)$.
    				}{
    					$\Delta_e \leftarrow \Delta_e - 1$ for $e = (v^*,u')$.
    				}
    			}
    	    }
		}
		\ForEach{$\{v^*,v^-\} \in \{\{v,v'\},\{v',v\}\}$}{
    		\ForEach{$\{u,u'\} \subset N_{v^*} \backslash N_{v^-}$}{
    			\If{$(u,u') \notin E$}{
    				%$E_c = \{(u,u'),(v^-,u),(v^-,u')\}$.\\
    				$\Delta_e \leftarrow \Delta_e + 1$ for $e \in \{(u,u'),(v^-,u),(v^-,u')\}$\\%E_c$.\\
    				$\Delta_e^* \leftarrow \Delta_e^* + 1$ for $e \in \{(u,u'),(v^-,u),(v^-,u')\}$\\%E_c$.
    			}
    		}
        }
	}
	Output $G_c = (V,E)$.
\end{algorithm2e}

This concludes the method of introducing additional conflicts for a polynomial-time near-optimal scheduling. 
%METHODOLOGY

%\sub
\section{Method II - Physical Modifications in Network}
\label{subsec:phys}

There are another possibilities for getting claw-freeness. The idea is to propose some physical modifications which include position adjustment, transmission range adjustment and antenna orientation adjustment in the network configuration. These possible interventions may lead to changes in the set of possible transmissions and interference between them. By making use of this property, our goal is to achieve claw-freeness. In this type of modification, we make changes on the network which automatically lead to some changes in the conflict graph. 

\vspace{-0.5cm}
\begin{figure}[H]
  \centering
    \includegraphics[width=0.42\textwidth]{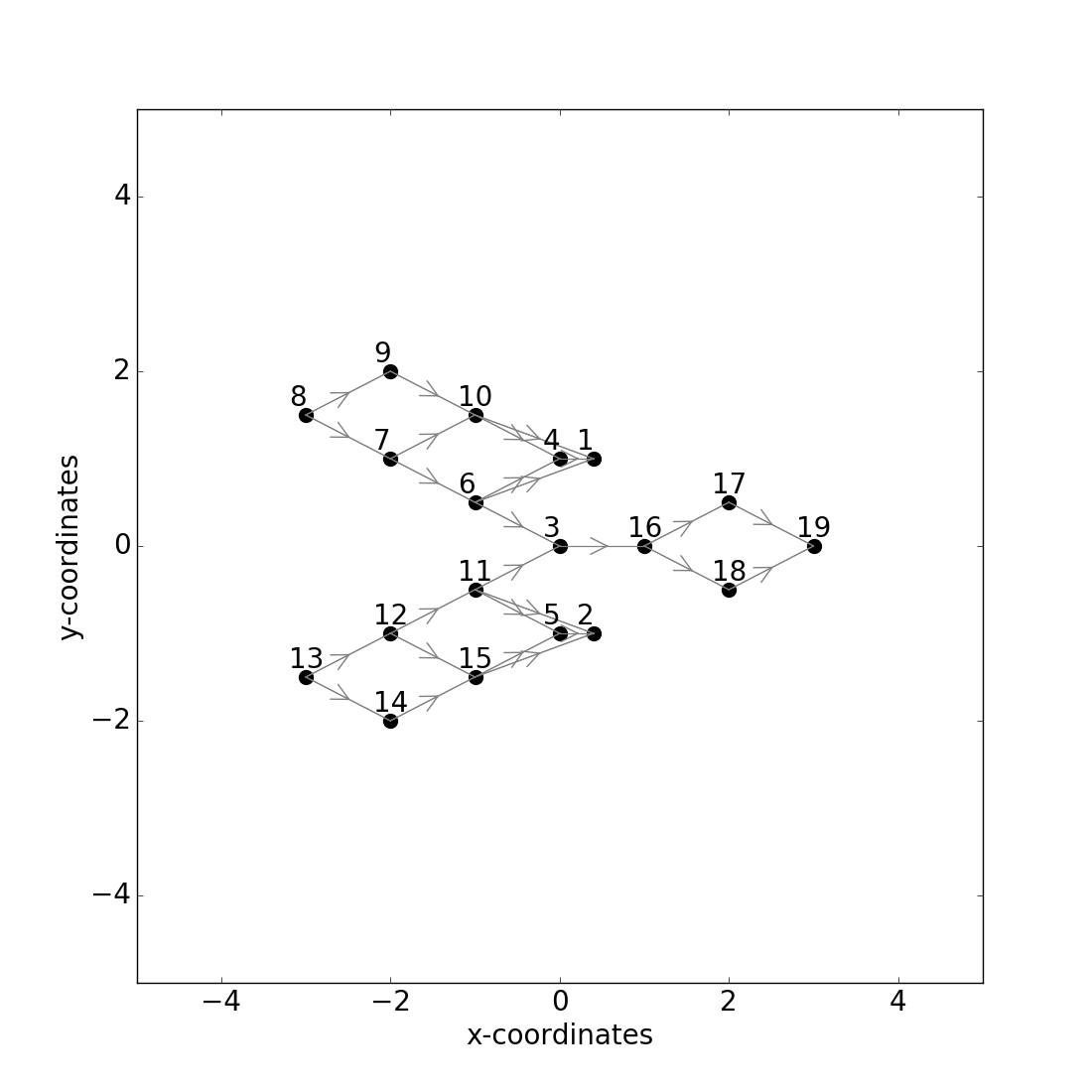}    
    \vspace{-0.5cm}
    \caption{An example network for illustration of physical modifications.}
    \label{fig:guici6}
\end{figure}
\vspace{-0.75cm}

\begin{figure}[h]
  \centering
    \includegraphics[width=0.45\textwidth]{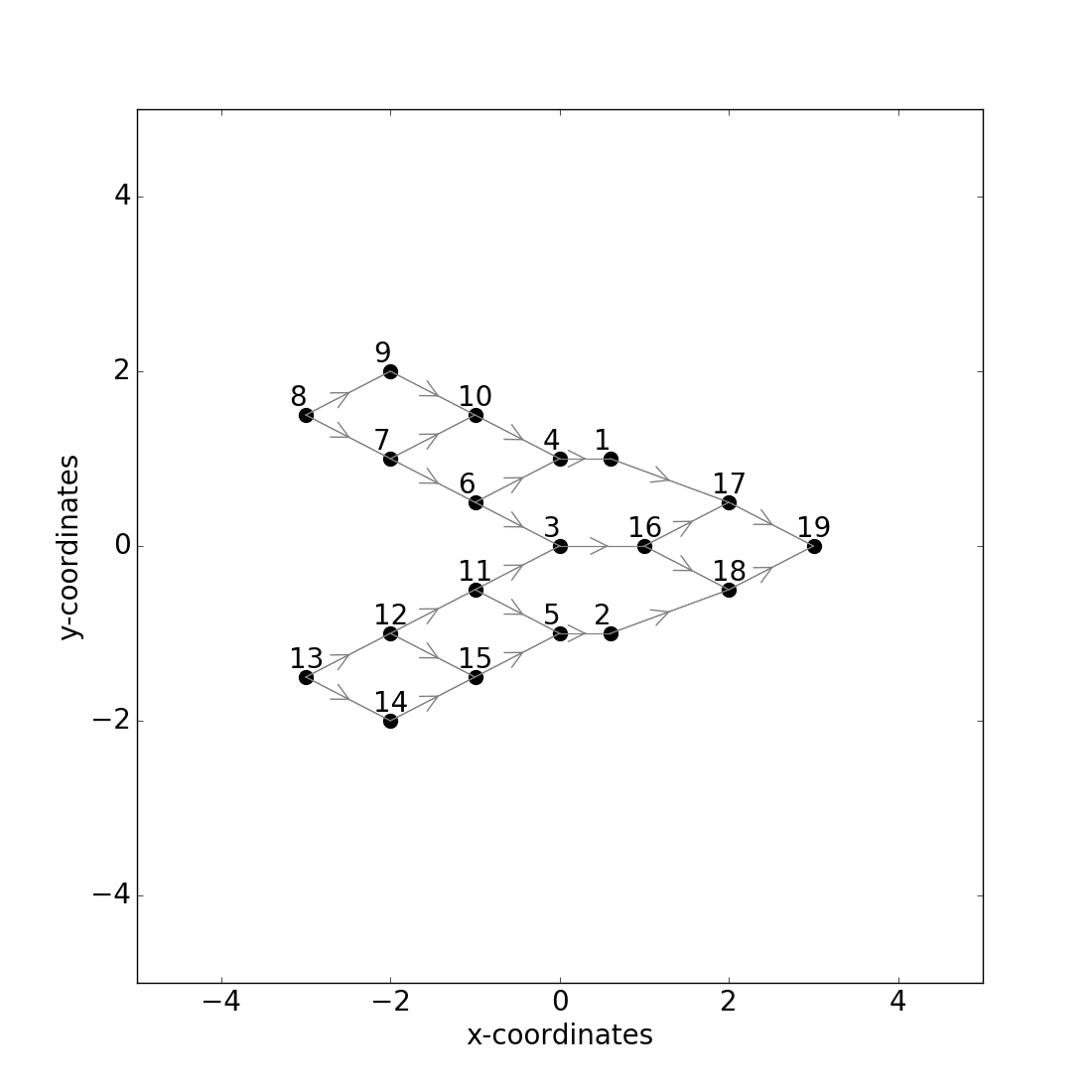}
    \vspace{-0.5cm}
    \caption{Network seen in Fig. \ref{fig:guici6} is modified using position adjustment.}
    \label{fig:guici7}
    \vspace{-0.75cm}
\end{figure}

\textbf{Position adjustment:} Let us consider the network given in Fig. \ref{fig:guici6}. We observe transceivers as the black dots and possible transmissions as directed gray edges. Transmission range $r_T$ is assumed to be equal for all transceivers. Considered network has 32 claws in its conflict graph with current connections. Also, if the capacity of link between the transceivers 3 and 16 is very low, this link serves as the bottleneck in the network and it can drastically decrease the end-to-end throughput in the system. By a minor change in the transceiver locations in 2-D coordinate system, $x_1=x_1+0.2$ and $x_2=x_2+0.2$ where $x_i$ is the $x$-coordinate of transceiver $i$ in figure, we reach a configuration seen in Fig. \ref{fig:guici7}. By shifting the locations of two transceivers, we lose transmissions $(6,1)$, $(10,1)$, $(6,\{1,4\})$, $(10,{1,4})$, $(11,2)$, $(15,2)$, $(11,\{2,5\})$, $(15,\{2,5\})$, whereas $(1,17)$ and $(2,18)$ show up as new possible transmissions in the conflict graph. This new configuration has a claw-free conflict graph and also solves the bottleneck problem seen in Fig. \ref{fig:guici6} by introducing 2 new paths for the data exchange between clusters of transceivers. Using this example, we conclude that it is possible to get a claw-free conflict graph to do scheduling in polynomial time while increasing the throughput in the network by adjusting the positions of transceivers.

\vspace{-0cm}

\textbf{Transmission range adjustment:} Let us again consider the network given in Fig. \ref{fig:guici6}. Instead of assuming equal transmission range of $r_T$ for all transceivers, suppose we can assign a different transmission range for each transceiver. If we increase the transmission range of transceivers 1 and 2 and decrease the ones of 6, 10, 11, 15, the conflict graph of the network becomes claw free. In this case, position adjustment and transmission range adjustment is equivalent in terms of conflict graph modifications because they lead to same connection setup in the network. Nevertheless, it may not be wise to use the Protocol model if transmission range of transceivers can differ.

Making physical modifications in a network is not feasible if there are strict requirements on the locations of nodes or there is no possibility or resources to make such changes. Even if this was not an issue, we are not able to offer an automatized algorithm to do these modifications.

\section{Mixed Scheduling Strategy}
\label{mixedsc}

We are able to detect the locations of transceivers which are included in the transmissions resulting to be a node belonging to a claw in the conflict graph. Such an example of a heat map can be seen in Fig. \ref{heat_map}. Assume that all of the claws are stemmed from the transceivers that are located in a specific part, Partition I, of the network and remaining part, Partition II, consists of the transceivers such that the possible transmissions of these transceivers do not lead to any claws in the conflict graph. For such networks, we propose a mixed scheduling approach. By doing so, we do not intervene neither physical structure of network nor conflict graph, and also do not lose the advantage of claw-freeness coming from the remaining part. Strategy can be seen in Algortihm \ref{msa}. The algorithm works in a divide and conquer fashion since we divide the main problem into subproblems, solve these subproblems and combine the solutions in the end. This proposed scheme exploits the advantage of claw-freeness of a part of the conflict graph even if the complete conflict graph is not claw-free. 

\begin{comment}
The proposed algorithm first constructs the conflict graph of the overall network using possible transmissions and conflicts between them. Then, we divide this overall conflict graph into two where one is for the Partition I where claws come from and second is for the claw-free part, Partition II. While doing this, transmissions containing transceivers from both parts are assigned to Partition I's conflict graph. In the next step, we find MWIS in the conflict graph having claws using Traskov et al.'s \cite{traskov2012scheduling} nearly optimal algorithm and find MWIS in the claw-free conflict graph using Minty's \cite{minty1980maximal,nakamura2001revision} or Faenza et al.'s \cite{faenza2014solving} optimal algorithm. 
\end{comment}

In the last part of the algorithm, since there may be some nodes in the general independent set that are connected in the original conflict graph $\mathcal{G}=(\mathcal{V}, \mathcal{E})$ of the overall network, we exclude the ones having smaller weight in such pairs. 

\begin{comment}
In other words, consider two possible transmissions $S_k$ and $S_l$ where $\{S_k,S_l\} \in  \mathcal{E}$ in $\mathcal{G}=(\mathcal{V}, \mathcal{E})$ and $S_k\in I_1$, $S_l\in I_2$ where $I_1$ and $I_2$ are the resulting independent sets coming from two partitions. In this situation, we have to cancel less weighted one of $S_k$ and $S_l$.
\end{comment}

\vspace{0.5cm}

\begin{algorithm2e}[H]
%{\LinesNumberedHidden
%    \begin{algorithm}[H]
    \small
        \SetKwInOut{Input}{Input}
        \SetKwInOut{Output}{Output}
        %\SetAlgorithmName{Algorithm}{}

        \Input{$\mathcal{T}_1$ --> Transceiver set in    Partition I. \\ $\mathcal{T}_2$ --> Transceiver set in Partition II.\\ $A_1$ --> An approximation algorithm for finding MWIS in general graphs.\\ $A_2$ --> A precise algorithm for finding MWIS in claw-free graphs.}
         
        \Output{$I$ --> MWIS of $\mathcal{G}=(\mathcal{V}, \mathcal{E})$} 
        \BlankLine
        
        Construct the conflict graph $\mathcal{G}=(\mathcal{V}, \mathcal{E})$ of the given network where possible transmissions are denoted as $S_k=(i_k,J_k)$, $S_k\in \mathcal{V}$.

        Divide the conflict graph $\mathcal{G}=(\mathcal{V}, \mathcal{E})$ into two disjoint conflict graphs $\mathcal{G}_1=(\mathcal{V}_1, \mathcal{E}_1)$  and $\mathcal{G}_2=(\mathcal{V}_2, \mathcal{E}_2)$ such that:
        
        \For{$k=1$ to $|V|$}{
        
        \If{ ($i_k\in \mathcal{T}_1$) $||$  ($j\in \mathcal{T}_1$ for $\exists j\in J_k$)}{
            $S_k\in \mathcal{V}_1$
        }
        
        \Else{
            $S_k\in \mathcal{V}_2$
        }

        }
        
        $I_1 = A_1(G_1)$
        
        $I_2 = A_2(G_2)$
        
        $I = I_1 \cup I_2$
        
        Delete the elements of $I$ such that:
        
        \For{$k=1$ to $length(I)$}{
        \For{$l=k+1$ to $length(I)$}{
        
        \If{$\{I[k],I[l]\} \in  \mathcal{E}$ in $\mathcal{G}$}{
            del-min-weighted($I[k],I[l]$).
        }
        
        \Else{
            Do nothing.
        }

        }
        }

\caption{Mixed Scheduling Algorithm}
\label{msa}
\end{algorithm2e}

\section{Simulation Results and Discussion}
\label{simulasyon}

To evaluate the performance of the claw breaking strategy, we conduct simulations over randomly located $n$ transceivers in a 2-D coordinate system with an area of 20x20. Simulation results give us the weights of MWIS. Since, weight of the resulting independent set and network throughput are proportional, we consider MWIS weight as maximum network throughput in figures. We compare the performance of claw breaking with the optimal performance and with the maximal set scheduling method. To measure the optimal performance, namely MWIS, is NP-hard, but exploiting the fact that MWIS cannot include two different transmission nodes such that $i_1=i_2$, we are able to find the MWIS of the original conflict graph, which contains claws, in a reasonable time for the networks having not so many transceivers. Our goal is to see how suboptimal our strategy is. Weights of the possible transmission nodes are given according to the cardinality of their receiver set $J$. Our assumptions on network can be seen in Section \ref{sec:model2}.

\begin{figure}[h]
  \centering
    \includegraphics[width=0.43\textwidth]{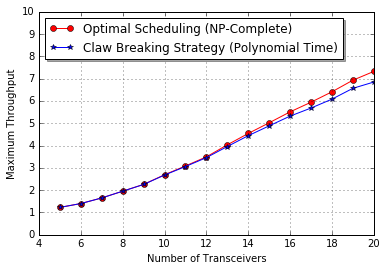}    
    \vspace{-0.5cm}
    \caption{Performance analysis of Claw Breaking Strategy with respect to number of transceivers in the network. Assumptions can be seen in Section \ref{simulasyon} and $r_T=7$.}
    \label{numbernode}
\end{figure}

We can see the change of MWIS with respect to the number of transceivers in the network in Fig. \ref{numbernode}. For $n<15$, claw breaking performs optimally, actually this is because we do not observe high number of claws for low number of transceivers. As a reminder, we have 0.19 claws in average for $n=10$ as seen in TABLE \ref{teybil2}. But, even for $n=20$, where we have a denser network and observe higher number of claws, our strategy performs nearly optimal, 93\% of optimal performance.

\vspace{-0.75cm}

\begin{figure}[t] % "[t!]" placement specifier just for this example
\begin{subfigure}{0.43\textwidth}
\includegraphics[width=\linewidth]{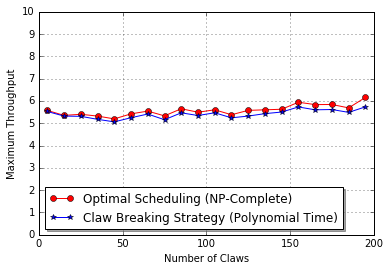}
\caption{$n=10$ and $r_T=random(10,11,12,13)$} \label{10nodevsclawperformance}
\end{subfigure}\hspace*{\fill}
\begin{subfigure}{0.43\textwidth}
\includegraphics[width=\linewidth]{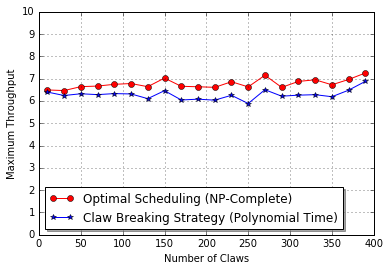}
\caption{$n=15$ and $r_T=random(8,9,10,11)$} \label{15nodevsclawperformance}
\end{subfigure}

\vspace{-0.5cm}
\medskip

\begin{subfigure}{0.43\textwidth}
\includegraphics[width=\linewidth]{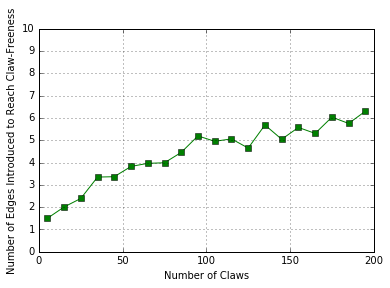}
\caption{$n=10$ and $r_T=random(10,11,12,13)$} \label{10nodevsclawedge}
\end{subfigure}\hspace*{\fill}
\begin{subfigure}{0.43\textwidth}
\includegraphics[width=\linewidth]{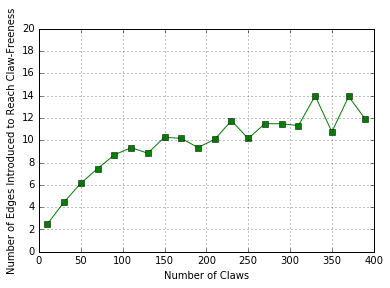}
\caption{$n=15$ and $r_T=random(8,9,10,11)$} \label{15nodevsclawedge}
\end{subfigure}

\vspace{-0.5cm}
\caption{Performance analysis of Claw Breaking Strategy under the assumptions given in Section \ref{simulasyon}.}
\label{fig:11111}
\vspace{-1cm}
\end{figure}

Then, we fix the number of transceivers to $n=10$, and we make random choices between $r_T=10,11,12,13$ in every iteration to get enough samples of conflict graphs having claws between 1 and 200. Performance evaluation with respect to number of claws in the conflict graph can be seen in Fig. \ref{10nodevsclawperformance}. Even though increasing number of claws decreases the performance of our strategy, this decrease is very small and our strategy performs nearly optimal. Average number of edges introduced to reach claw-freeness can be seen in Fig. \ref{10nodevsclawedge}. Connected networks tend to have higher number of claws. Nevertheless, since we are able to break hundreds of claws by just introducing a few edges in the conflict graph, our strategy also performs well for connected networks.

As a next step, we increase the number of transceivers and fix it to $n=15$, and make random choice between $r_T=8,9,10,11$ in every iteration. This time, we evaluate the cases having claws until 400. Above this value, we do not get enough samples since networks in this setup do not tend to have more than 400 claws in their conflict graphs. Performance of our strategy can be seen in Fig. \ref{15nodevsclawperformance}. We observe that the margin between claw breaking and optimal performance is higher than the case in Fig. \ref{10nodevsclawperformance}, but we do not even lose one transmission in average, so claw breaking strategy performs nearly optimal for $n=15$, 88\% for the worst sample point (conflict graphs having claws between 245-255). On the other hand, connectedness ratio is significantly lower in this case. This is an expected result since we have higher number of transceivers and therefore we decrease $r_T$ not to have unnecessarily high number of possible transmissions in the system. Decrease of connectedness ratio is due to randomness of transceiver locations and can be compensated by a different placing method, for instance: random placement on a grid. Number of edges introduced in order to get rid of claws can be seen in Fig. \ref{15nodevsclawedge}. We observe from Fig. \ref{10nodevsclawedge} and Fig. \ref{15nodevsclawedge} that introduced number of edges for claw-freeness behaves as a concave function. This observation is a very motivating factor of this strategy to be used in networks having very high number of claws, say thousands, and being nearly optimal.

\begin{figure}[H] 
\begin{subfigure}{0.43\textwidth}
\includegraphics[width=\linewidth]{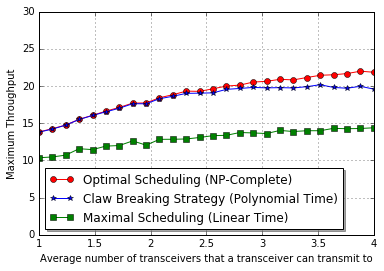}
\caption{Performance comparison of Claw Breaking Strategy and Maximal Set Scheduling. $n=32$.} \label{32node}
\end{subfigure}\hspace*{\fill}
\begin{subfigure}{0.43\textwidth}
\includegraphics[width=\linewidth]{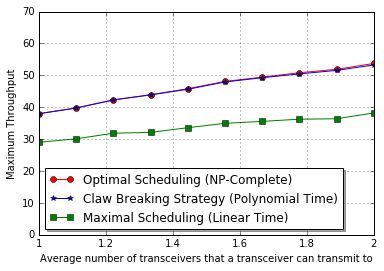}
\caption{Performance comparison of Claw Breaking Strategy and Maximal Set Scheduling. $n=100$.} \label{100node}
\end{subfigure}
\vspace{-0.5cm}
\caption{Performance comparison of Claw Breaking Strategy with Maximal Set Scheduling under the assumptions given in Section \ref{simulasyon}.} \label{fig:22222}
\end{figure}

Lastly, we have a set of simulations for higher transceiver numbers as seen from Fig. \ref{32node} and \ref{100node}. In these simulations, we evaluate and compare the performances with respect to average number of neighbors that a transceiver has. For both cases, until 2 neighbors, our strategy performs optimal and 25-50\% better than the maximal set scheduling. We can observe that the performance of claw breaking strategy starts to decrease after average 2.5 neighbors but still performs 88\% of the optimal result when we reach 4 neighbors in average. Note that, claw breaking performs 33\% better than the maximal set scheduling even in worst case as seen in Fig. \ref{32node}.

We have two main limitations in claw breaking strategy. First, we have to limit the number of neighbors of a transceiver, otherwise we have an exponential complexity to construct the conflict graph as explained in Section \ref{sc1}. Second, directed antennas should be deployed in the network setup, if not, we may observe a very high number of claws even for relatively low number of transceivers. For the second limitation, we can still implement our method, but it does not perform as good as in case of directed antennas. In general, network should not be very dense in terms of possible connections.

\vspace{-1cm}

\section{Conclusion}
\label{concl}

In this paper, we address the scheduling problem in wireless ad hoc networks. We set up some networks which have claw-free conflict graphs under various assumptions. These networks can be throughput optimally scheduled in polynomial time. A major drawback is that most of the real life networks do not have claw-free conflict graphs therefore throughput optimal scheduling is limited to a small number of networks.

To address for the scheduling of more general networks, we offer two different approaches. First, we can break the claws on the conflict graph by introducing edges without any intervention to network setup. This is a suboptimal approach but after proper optimization on the selection of new edges the decrease in throughput can be kept minimal. Second approach is to make physical modifications in considered network to change connections and/or interference relationships to get a claw-free conflict graph. Such physical modifications include position adjustment, transmission range adjustment and antenna orientation adjustment of transceivers. Disadvantage of physical modifications is that we are not able to give an automatized method to implement them with minimum intervention to network setup. Therefore, we need a human expert to offer such modifications to reach claw-freeness in this second approach.

We propose a different approach to scheduling where claw-free zones and zones introducing claws are scheduled differently. Zones with claws are scheduled with an approximate scheduling algorithm whereas the rest of the network is scheduled using a throughput optimal polynomial time algorithm. Then, resulting independent sets are carefully merged. Mixed scheduling algorithm makes polynomial time scheduling possible for the parts of the network which induce claw-free conflict graphs. It is superior to approximate scheduling algorithms in that regard. For this method to be applicable, claws in the conflict graph should come from a specific part of the network.

From the simulations, we observe that claw breaking strategy works nearly optimal for various number of transceivers, up to a limited number of connections. The only limitation is the need for directed antennas. Because, with omnidirectional antennas, the number of claws in the conflict graph becomes very high and shows a very rapid increase with the transmission range of transceivers, corresponding to the number of receivers they connect. Also, deployment of omnidirectional antennas almost always increases the number of neighbors a transceiver has, therefore increasing the tendency of a network to break the rule $|N(i_x)|\leq K$. Thus, deployment of directed antennas is crucial for both construction of the conflict graph and for the better performance of claw breaking strategy.

\ifCLASSOPTIONcaptionsoff
  \newpage
\fi

\bibliographystyle{IEEEtran}  
\bibliography{references} 

\begin{thebibliography}{10}
\providecommand{\url}[1]{#1}
\csname url@samestyle\endcsname
\providecommand{\newblock}{\relax}
\providecommand{\bibinfo}[2]{#2}
\providecommand{\BIBentrySTDinterwordspacing}{\spaceskip=0pt\relax}
\providecommand{\BIBentryALTinterwordstretchfactor}{4}
\providecommand{\BIBentryALTinterwordspacing}{\spaceskip=\fontdimen2\font plus
\BIBentryALTinterwordstretchfactor\fontdimen3\font minus
  \fontdimen4\font\relax}
\providecommand{\BIBforeignlanguage}[2]{{%
\expandafter\ifx\csname l@#1\endcsname\relax
\typeout{** WARNING: IEEEtran.bst: No hyphenation pattern has been}%
\typeout{** loaded for the language `#1'. Using the pattern for}%
\typeout{** the default language instead.}%
\else
\language=\csname l@#1\endcsname
\fi
#2}}
\providecommand{\BIBdecl}{\relax}
\BIBdecl

\bibitem{arikan1984some}
E.~Arikan, ``Some complexity results about packet radio networks (corresp.),''
  \emph{IEEE Transactions on Information Theory}, vol.~30, no.~4, pp. 681--685,
  1984.

\bibitem{ephremides1990scheduling}
A.~Ephremides and T.~V. Truong, ``Scheduling broadcasts in multihop radio
  networks,'' \emph{IEEE Transactions on communications}, vol.~38, no.~4, pp.
  456--460, 1990.

\bibitem{sharma2006complexity}
G.~Sharma, R.~R. Mazumdar, and N.~B. Shroff, ``On the complexity of scheduling
  in wireless networks,'' in \emph{Proceedings of the 12th annual international
  conference on Mobile computing and networking}.\hskip 1em plus 0.5em minus
  0.4em\relax ACM, 2006, pp. 227--238.

\bibitem{hajek1988link}
B.~Hajek and G.~Sasaki, ``Link scheduling in polynomial time,'' \emph{IEEE
  transactions on Information Theory}, vol.~34, no.~5, pp. 910--917, 1988.

\bibitem{traskov2012scheduling}
D.~Traskov, M.~Heindlmaier, M.~M{\'e}dard, and R.~Koetter, ``Scheduling for
  network-coded multicast,'' \emph{IEEE/ACM Transactions on Networking (TON)},
  vol.~20, no.~5, pp. 1479--1488, 2012.

\bibitem{bao2001new}
L.~Bao and J.~Garcia-Luna-Aceves, ``A new approach to channel access scheduling
  for ad hoc networks,'' in \emph{Proceedings of the 7th annual international
  conference on Mobile computing and networking}.\hskip 1em plus 0.5em minus
  0.4em\relax ACM, 2001, pp. 210--221.

\bibitem{gupta2000capacity}
P.~Gupta and P.~R. Kumar, ``The capacity of wireless networks,'' \emph{IEEE
  Transactions on information theory}, vol.~46, no.~2, pp. 388--404, 2000.

\bibitem{ahlswede2000network}
R.~Ahlswede, N.~Cai, S.-Y. Li, and R.~W. Yeung, ``Network information flow,''
  \emph{IEEE Transactions on information theory}, vol.~46, no.~4, pp.
  1204--1216, 2000.

\bibitem{ho2003randomized}
T.~Ho, M.~Medard, J.~Shi, M.~Effros, and D.~R. Karger, ``On randomized network
  coding,'' in \emph{Proceedings of the Annual Allerton Conference on
  Communication Control and Computing}, vol.~41, no.~1.\hskip 1em plus 0.5em
  minus 0.4em\relax The University; 1998, 2003, pp. 11--20.

\bibitem{ho2006random}
T.~Ho, M.~M{\'e}dard, R.~Koetter, D.~R. Karger, M.~Effros, J.~Shi, and
  B.~Leong, ``A random linear network coding approach to multicast,''
  \emph{IEEE Transactions on Information Theory}, vol.~52, no.~10, pp.
  4413--4430, 2006.

\bibitem{tassiulas1992stability}
L.~Tassiulas and A.~Ephremides, ``Stability properties of constrained queueing
  systems and scheduling policies for maximum throughput in multihop radio
  networks,'' \emph{IEEE transactions on automatic control}, vol.~37, no.~12,
  pp. 1936--1948, 1992.

\bibitem{sundararajan2006systematic}
J.~K. Sundararajan, M.~M{\'e}dard, R.~Koetter, and E.~Erez, ``A systematic
  approach to network coding problems using conflict graphs,'' in
  \emph{Proceedings of the UCSD Workshop on Information Theory and its
  Applications}, 2006.

\bibitem{sundararajan2007network}
J.~K. Sundararajan, M.~M{\'e}dard, M.~Kim, A.~Eryilmaz, D.~Shah, and
  R.~Koetter, ``Network coding in a multicast switch,'' in \emph{INFOCOM 2007.
  26th IEEE International Conference on Computer Communications. IEEE}.\hskip
  1em plus 0.5em minus 0.4em\relax IEEE, 2007, pp. 1145--1153.

\bibitem{minty1980maximal}
G.~J. Minty, ``On maximal independent sets of vertices in claw-free graphs,''
  \emph{Journal of Combinatorial Theory, Series B}, vol.~28, no.~3, pp.
  284--304, 1980.

\bibitem{nakamura2001revision}
D.~Nakamura and A.~Tamura, ``A revision of minty's algorithm for finding a
  maximum weight stable set of a claw-free graph,'' \emph{Journal of the
  Operations Research Society of Japan}, vol.~44, no.~2, pp. 194--204, 2001.

\bibitem{schrijver2003combinatorial}
A.~Schrijver, \emph{Combinatorial optimization: polyhedra and
  efficiency}.\hskip 1em plus 0.5em minus 0.4em\relax Springer Science \&
  Business Media, 2003, vol.~24.

\bibitem{faenza2014solving}
Y.~Faenza, G.~Oriolo, and G.~Stauffer, ``Solving the weighted stable set
  problem in claw-free graphs via decomposition,'' \emph{Journal of the ACM
  (JACM)}, vol.~61, no.~4, p.~20, 2014.

\bibitem{2017arXiv171101620K}
A.~{Kose} and M.~{Medard}, ``{Scheduling Wireless Ad Hoc Networks in Polynomial
  Time Using Claw-free Conflict Graphs},'' \emph{ArXiv e-prints}, Nov. 2017.

\bibitem{chudnovsky2005structure}
M.~Chudnovsky and P.~D. Seymour, ``The structure of claw-free graphs.''
  \emph{Surveys in combinatorics}, vol. 327, pp. 153--171, 2005.

\end{thebibliography}

\end{document}